\documentclass[twocolumn,aps,pra,floatfix,superscriptaddress]{revtex4-1}
\usepackage{amsmath,amssymb}
\usepackage[colorinlistoftodos, color=green!40, prependcaption]{todonotes}
\usepackage{graphicx,float}
\usepackage{times,txfonts}
\usepackage{textcomp}
\usepackage{color}
\usepackage{multirow}
\usepackage{color}
\usepackage{soul}
\usepackage{hyperref}
\usepackage{natbib}
\usepackage{nicefrac}
\usepackage{multirow}
\usepackage{blindtext}
\usepackage[export]{adjustbox}
\usepackage[T1]{fontenc}

\begin{document}
\title{Single-photon detectors on arbitrary photonic substrates}

\author{Max~Tao}
\thanks{These two authors contributed equally}
\affiliation{Research Laboratory of Electronics, Massachusetts Institute of Technology, Cambridge, Massachusetts 02139, USA}

\author{Hugo~Larocque}
\thanks{These two authors contributed equally}
\affiliation{Research Laboratory of Electronics, Massachusetts Institute of Technology, Cambridge, Massachusetts 02139, USA}

\author{Samuel~Gyger}
\affiliation{Research Laboratory of Electronics, Massachusetts Institute of Technology, Cambridge, Massachusetts 02139, USA}
\affiliation{KTH Royal Institute of Technology, Stockholm, Sweden}

\author{Marco~Colangelo}
\affiliation{Research Laboratory of Electronics, Massachusetts Institute of Technology, Cambridge, Massachusetts 02139, USA}

\author{Owen~Medeiros}
\affiliation{Research Laboratory of Electronics, Massachusetts Institute of Technology, Cambridge, Massachusetts 02139, USA}

\author{Ian~Christen}
\affiliation{Research Laboratory of Electronics, Massachusetts Institute of Technology, Cambridge, Massachusetts 02139, USA}

\author{Hamed~Sattari}
\affiliation{Centre Suisse d’Electronique et de Microtechnique, Neuchatel, Switzerland}

\author{Gregory~Choong}
\affiliation{Centre Suisse d’Electronique et de Microtechnique, Neuchatel, Switzerland}

\author{Yves~Petremand}
\affiliation{Centre Suisse d’Electronique et de Microtechnique, Neuchatel, Switzerland}

\author{Ivan~Prieto}
\affiliation{Centre Suisse d’Electronique et de Microtechnique, Neuchatel, Switzerland}

\author{Yang~Yu}
\affiliation{Raith America Inc., Troy, NY, USA}

\author{Stephan~Steinhauer}
\affiliation{KTH Royal Institute of Technology, Stockholm, Sweden}

\author{Gerald~L.~Leake}
\affiliation{State University of New York Polytechnic Institute, Albany, New York 12203, USA}

\author{Daniel~J.~Coleman}
\affiliation{State University of New York Polytechnic Institute, Albany, New York 12203, USA}

\author{Amir~H.~Ghadimi}
\affiliation{Centre Suisse d’Electronique et de Microtechnique, Neuchatel, Switzerland}

\author{Michael~L.~Fanto}
\affiliation{Air Force Research Laboratory, Information Directorate, Rome, New York, 13441, USA}

\author{Val~Zwiller}
\affiliation{KTH Royal Institute of Technology, Stockholm, Sweden}

\author{Dirk~Englund}
\affiliation{Research Laboratory of Electronics, Massachusetts Institute of Technology, Cambridge, Massachusetts 02139, USA}

\author{Carlos~Errando-Herranz}
\email{c.errandoherranz@tudelft.nl}
\affiliation{Research Laboratory of Electronics, Massachusetts Institute of Technology, Cambridge, Massachusetts 02139, USA}
\affiliation{Institute of Physics, University of M\"unster, 48149, M\"unster, Germany}
\affiliation{Department of Quantum and Computer Engineering, Delft University of Technology, Delft, Netherlands}
\affiliation{QuTech, Delft University of Technology, Delft, Netherlands}


\begin{abstract}
Detecting non-classical light is a central requirement for photonics-based quantum technologies. Unrivaled high efficiencies and low dark counts have positioned superconducting nanowire single photon detectors (SNSPDs) as the leading detector technology for fiber and integrated photonic applications. However, a central challenge lies in their integration within photonic integrated circuits regardless of material platform or surface topography. Here, we introduce a method based on transfer printing that overcomes these constraints and allows for the integration of SNSPDs onto arbitrary photonic substrates. We prove this by integrating SNSPDs and showing through-waveguide single-photon detection in commercially manufactured silicon and lithium niobate on insulator integrated photonic circuits. Our method eliminates bottlenecks to the integration of high-quality single-photon detectors, turning them into a versatile and accessible building block for scalable quantum information processing. 
\end{abstract}

\maketitle

\section{Introduction}
\begin{figure*}[htbp]
  \centering
  \includegraphics[width=\linewidth]{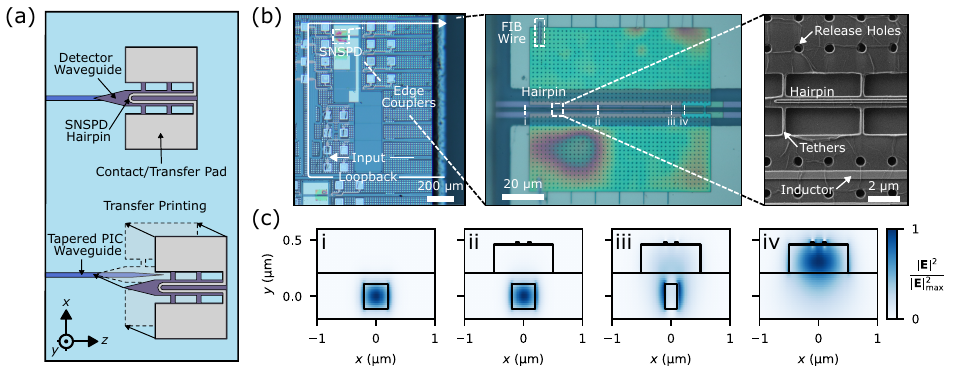}
\caption{\textbf{Hybrid integrated SNSPD assembly and modeling} (a) Schematics of the hybrid integration process of SNSPDs on foundry PICs. (b) Optical micrograph of a foundry silicon PIC with integrated SNSPDs. Insets: optical micrograph of an assembled device, and a scanning electron micrograph of the hairpin detector. (c) Fundamental mode profile of the hybrid mode converter at distances of (i) 0~\textmu m (ii) 56~\textmu m (iii) 96~\textmu m and (iv) 100~\textmu m from the start of the structure's detector waveguide.}
\label{fig:fig1}
\end{figure*}

Optical quantum technologies are central to quantum computing~\cite{Ladd:10}, communication~\cite{Gisin:07}, and simulation~\cite{Aspuru-guzik:12}. Scaling these technologies to the system sizes required by quantum applications has motivated the development of quantum photonic integrated circuits (PICs)~\cite{Obrien:09}, which leverage standardized semiconductor manufacturing for producing large-scale optical systems. A key requirement of such systems is the detection of single photons for tasks ranging from preparing and measuring quantum states to implementing quantum gates~\cite{Ferrari:18, Wang:20}. Superconducting nanowire single--photon detectors (SNSPDs) are among the best single--photon detectors available today due to their combination of record high detection efficiency~\cite{Reddy:20}, broadband operation~\cite{Colangelo:22}, low dark counts~\cite{Chiles:22}, fast recovery time~\cite{Munzberg:18}, very low timing uncertainty~\cite{Goltsman:01}, and compatibility with photonic reconfiguration~\cite{Gyger:21}.

A central challenge in the development of quantum PICs involves integrating SNSPDs within large-scale circuits with (i) sufficient fabrication yields and (ii) universal methods that seamlessly carry over within PIC platforms. Monolithic integration can readily provide sufficient yield levels, yet often involves process flows hyper-specialized to a particular fabrication node~\cite{Najafi:15ii,Cheng:19}. To address this issue, recent advances in hybrid quantum PICs~\cite{Elshaari:20, Kim:20} motivated the development of micrometer-scale flip chip processes for integrating SNSPDs on a wider range of PICs~\cite{Najafi:15}. However, successful flip chip transfers require meticulous handling of the SNSPDs with equipment such as tungsten microprobes. Furthermore, this method requires highly accurate structural features conforming with those of the SNSPD, not present in the vast majority of PIC platforms. Both of these drawbacks could prevent deploying SNSPD flip-chip transfers at scale, thereby warranting a hybrid integration method that simultaneously overcomes: (i) fabrication incompatibilities among PIC platforms, with a prominent example being lithium niobate on insulator (LNOI) which require customized SNSPD fabrication flows~\cite{Sayem:20,Lomonte:21,Colangelo:24}; (ii) limited device yields~\cite{Najafi:15ii, Colangelo:22}; and (iii) lack of control over the PIC fabrication process, which can be especially common while integrating with large-scale foundry-processed PICs~\cite{Hochberg:10,Fahrenkopf:19,Alexander:24}.

Here, we address these challenges via the hybrid integration of SNSPDs on PICs by transfer printing.
Our method relies on (i) standardized SNSPD fabrication~\cite{Gyger:21} to avoid incompatibilities with PIC fabrication, (ii) preliminary screening allowing us to selectively transfer functioning devices and overcome yield limitations in monolithic PIC platforms, and (iii) PIC structures compatible with arbitrary PICs that do not require flip chip bonding. 
We demonstrate the versatility of our method by integrating SNSPDs onto large-scale foundry silicon PICs and LNOI PICs, thereby confirming its compatibility with PIC platforms ranging from commercially manufactured systems to those that otherwise require substantially tailored SNSPD integration processes.

\section{Results}
Our device fabrication consists of (i) fabricating suspended silicon nitride waveguides topped with hairpin SNSPDs, (ii) detector screening via room-temperature resistance measurements, (iii) transfer printing, and (iv) wiring to ensure electrical connectivity.
The fabrication of our detectors draws on a process developed for MEMS-actuated PICs with SNSPDs~\cite{Gyger:21}. 
As outlined in Supplementary Section 1, the process produces suspended silicon nitride photonic waveguides topped with a NbTiN hairpin detector. Prior to transfer printing these devices, we measure their resistance at room temperature for screening purposes and select devices exhibiting a finite resistance for transfer. 

Drawing on established integration methods for on-chip light sources~\cite{Justice:12, Larocque:24}, we rely on a kinetically controlled elastomer stamp to transfer these SNSPDs on PICs. 
An optical microscopy apparatus loaded with a 0.41 numerical aperture objective allows us to monitor this transfer to preserve a sufficient level of alignment between the PIC and detectors. 
Figure~\ref{fig:fig1}(a) provides schematics of this integration process. 
The resulting devices feature hybrid optical mode converters formed by the PIC's native and the SNSPD's nitride waveguides, thereby enabling optical connectivity between the PIC and the transferred devices. To ensure electrical read-out from the SNSPD, we connect it to the PIC's electrical lines. 
This is done via in-situ focused-ion-beam (FIB) chemical vapor deposition of tungsten wires (see Fig.~\ref{fig:fig1}(b) and Supplementary Section 2 for more information). 

We then test the hybrid SNSPD-PIC structures in a closed loop cryostat with a base temperature of 0.78~K. 
Further details on the experimental setup are available in Supplementary Section 3. 
We confirm our detectors' ability to monitor photon counts by flood illuminating our chip before measuring their on-chip detection efficiency (ODE). 

We first integrate our SNSPDs on PICs commercially manufactured using a 193~nm deep-ultraviolet water-immersion lithography silicon photonic process. 
Figure~\ref{fig:fig1}(b) provides optical and scanning electron micrographs of the assembled structure. 
The resulting hybrid device adopts a mode converter inducing optical absorption into its SNSPD. 
As further elaborated in Supplementary Section 4, the PIC consists of single mode silicon waveguides operating at the O and C+L bands. FEM simulations suggest an optical absorption of 30.3\% at 1570 nm wavelengths However, we expect an observed angular offset between the PIC and detector waveguides to reduce this figure to 8.7\% .

We cryogenically test the hybrid SNSPD-silicon PIC device to confirm superconducting behavior, measuring a switching current of 7.1~\textmu A (see Supplementary Section 5 for the I-V curves).
We measure their ODE at C+L- and O-band telecommunication wavelengths compatible with the silicon waveguides of our PIC. 
We send 1570~nm and 1312~nm light from tunable external cavity diode lasers through a variable optical attenuator followed by a UHNA1 optical fiber array before going in the PIC by means of edge couplers at the chip's facet. 
We measure the optical transmission through the various stages of this fiber line (see Supplementary Table~1 for these pre-characterized values). 
We then confirm that the device can monitor counts by measuring a characteristic output pulse from the SNSPD under illumination through the PIC (see Supplementary Fig.~4b).

Figure~\ref{fig:countRate} plots the resulting photon and dark counts monitored by the detector, showing clear plateaus at both wavelengths that indicate high internal detection efficiency with low dark counts. We attribute the discrete and atypical appearance of our dark count data to the 100~ms integration time of our apparatus.
While biasing the detector with currents of 7~\textmu A and 6.4~\textmu A, we measured photon count rates of 1.356~MHz and 509~kHz for input wavelengths of 1570~nm and 1312~nm, respectively.
As shown in Supplementary Section 5, we confirm that these count-rates linearly drop with the intensity of our input light. 
The detector features a dark count rate of 40~Hz.
We compute the ODE by taking the ratio between the detector's photon counts and the photon flux, $\Phi$, in the silicon waveguide. 
Based on our apparatus, we define this metric as
\begin{equation}
    \label{eq:photonFlux}
    \Phi = \frac{1}{h\nu}\left[P_\text{in}\cdot 10^{-(\text{dB}_\text{f} + \text{dB}_\text{c} + \text{dB}_{\text{attn}})/10}\right],
\end{equation}
where dB$_\text{f}$ is the measured loss in dB through the fiber fed into the cryostat, dB$_\text{c}$ is the measured loss through the edge coupler, and dB$_\text{attn}$ is the setting of the variable attenuator. $P_\text{in}$ is the power supplied by the laser and $h\nu$ is the average input photon energy. 
As discussed in Supplementary Section 6, we measure dB$_\text{c}$ with an integrated loopback waveguide near the one coupled to the examined SNSPD. 
The extracted coupling efficiency value assumes identical optical coupling among the considered fiber-edge coupler pairs and also perfect alignment between the fiber array and the PIC.
Based on this estimate, our device showed a waveguide-coupled ODE of $7.8 \pm 0.2\%$ at 1570~nm and $7.1 \pm 0.1\%$ ODE at 1312~nm at bias currents of 7~\textmu A and 6.4~\textmu A, respectively (see Supplementary Section 7 for error calculations). 
These values share the same order of magnitude as our numerical estimates.

\begin{figure}[t]
  \centering
  \includegraphics[width=\columnwidth]{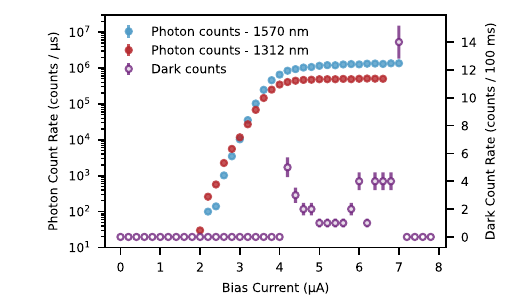}
\caption{\textbf{Detection efficiency} On-chip count rates acquired over 100 ms counting times of hybrid integrated SNSPDs for photons propagating through the waveguides of a silicon PIC.}
\label{fig:countRate}
\end{figure}

\begin{figure*}[t]
  \centering
\includegraphics[width=\linewidth]{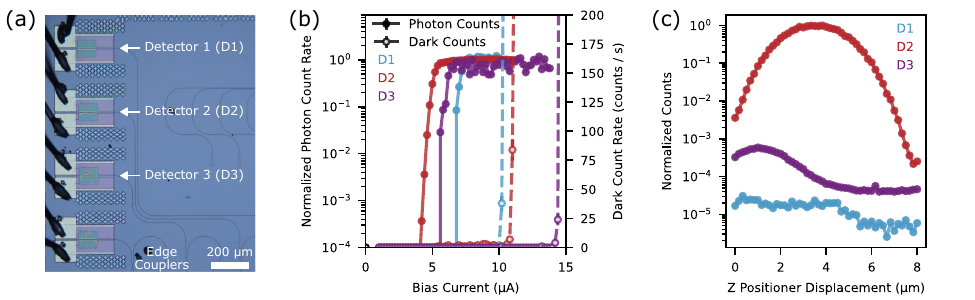}
\caption{\textbf{Integration of SNSPDs on LNOI} (a) LNOI PIC with hybrid integrated SNSPDs. (b) Photon and dark count rates of three SNSPDs on the LNOI chip measured at $650$~nm wavelength. (c) Relative count rates monitored by the transfered SNSPDs under fiber displacement demonstrate waveguide-coupled detection with 40~dB extinction.}
\label{fig:fig3}
\end{figure*}

We next illustrate the compatibility of our method with LNOI photonics. 
Direct SNSPD fabrication on such devices has proven challenging~\cite{Tanner:12} given the need for customized superconductor thin film deposition compatible with lithium niobate that avoids excessive substrate heating~\cite{Sayem:20,Lomonte:21,Colangelo:24}. 
Furthermore, adequate precautions must protect photonic waveguides during the detector fabrication~\cite{Sayem:20,Colangelo:24} or alternatively the detectors during the waveguide fabrication~\cite{Lomonte:21}. 
Our approach overcomes these issues by fabricating the detectors and waveguides on separate substrates, thereby motivating its integration in other PIC platforms facing similar fabrication challenges.
Figure~\ref{fig:fig3}(a) shows the transferred SNSPDs on the LNOI PICs. 
The LN waveguides consist of 200~nm thick straight waveguides surrounded by a 100~nm ridge. 
They have a width of 150~nm and a sidewall angle of $55^\text{o}$. 
We characterize these devices at optical wavelengths of 650~nm with the same methodology used for the silicon PICs, thereby  demonstrating efficiencies of up to $8.6 \pm 0.2$\% and low dark counts at saturation. 
We provide the corresponding photon and dark counts in Fig.~\ref{fig:fig3}(b), and measure a detector jitter of 242~ps as indicated in Supplementary Section 5.
The wide plateau suggests high internal detection efficiency.
We additionally monitor multiple on-chip detectors to gauge the relative influence between waveguide-coupled and stray photons on detected counts. 
As shown in Fig.~\ref{fig:fig3}(c), we couple light into the waveguide leading to one of the detectors, D2, while monitoring counts on all three detectors transferred on the PIC. 
Under optimal fiber to PIC coupling conditions, we observe a 40~dB extinction ratio between the coupled, D2, and uncoupled, D1 and D3, SNSPDs, or equivalently, between the device's waveguide-coupled and stray photons.
At lower fiber positions, we attribute the up to 10~dB higher counts for D1 and D3 to scattered light propagating through the buried oxide layer towards the detectors.

\section{Discussion}

As outlined in Supplementary Section 4, various approaches to co-designig the PIC and detector waveguides can increase detection efficiencies above our 7.8\% metric. For instance, switching the PIC waveguide to a longer silicon nitride one can increase the SNSPD's ODE to values near 99\%.
Furthermore, adopting measures such as distinct superconducting materials, better waveguide coupling, and reductions in kinetic inductance can lead to detectors with low jitter~\cite{Korzh:20}, compatibility with longer optical wavelengths~\cite{Pernice:12}, and shorter reset times~\cite{Munzberg:18}.
The angular alignment accuracy also affects mode conversion between the adiabatically tapered substrate and SNSPD waveguides (see Supplementary Section 6). 
Prior hybrid integration work of quantum photonic components on foundry PICs suggest that average offsets of 0.59$^\text{o}$ are statistically achievable under optimized transfer conditions~\cite{Larocque:24}. 
From simulations, we expect detectors with this offset to feature a similar ODE to that of a perfectly aligned device. 
Alternatively, relying on mode converter geometries that are more tolerant to misalignment could provide a path towards increasing the optical power transferred to the hybrid SNSPD~\cite{Bandyopadhyay:21}.

As in the case of prior hybrid approaches to SNSPD integration, ours allows for screening faulty devices to overcome device yield limitations. 
In addition, our use of elastomer stamps and detector-to-chip wiring gives us access to the full scaling potential of transfer-printing based technologies~\cite{Meitl:06,Justice:12}, which, in industrial settings, can reach transfer rates of up to two billion devices per hour with microscope-limited alignment tolerances~\cite{xCeleprint:18}.
Though stamp-based printing provides a means for a high-rate and automated detector transfer, their integration still requires the deposition of electrical contact lines. 
Our tungsten FIB-CVD deposition method conveniently wires the chip to the detectors without damaging them, yet it can be time-intensive. 
Alternative methods of electrically interfacing detectors, such as optical lithography followed by metal deposition or high precision printing of silver nano-inks can reliably produce our required low resistivity and micron-scale contacts~\cite{Lysien:22}. 

In summary, we demonstrated a hybrid integration method for interfacing SNSPDs with arbitrary photonic substrates. 
Using transfer printing and FIB-CVD on silicon PICs, we attain detector efficiencies of 7.8\% in the C+L-band and 7.1\% in the O-band along with dark counts of less than 100~Hz. 
In addition, we transferred devices onto LNOI chips and observed an 8.6\% efficiency, thereby demonstrating the versatility of our technique in regards to the PIC's material platform. 
Our results underscore SNSPD integration onto arbitrary PICs ranging from those manufactured at scale in a commercial foundry~\cite{Hochberg:10,Fahrenkopf:19,Alexander:24,Larocque:24} to those where monolithic detector fabrication can compromise the integrity of the PIC~\cite{Sayem:20,Lomonte:21,Colangelo:24}.
Enabling accessible integration of state-of-the-art single-photon detectors onto scalably manufacturable PICs implemented in arbitrary material platforms opens the door to fully-integrated quantum technologies for applications ranging from quantum communications with quantum repeaters~\cite{Ruf:21} to measurement-based quantum computing~\cite{Kok:07}, quantum sensing~\cite{Slussarenko:17}, and computing with trapped ions~\cite{Reens:22}.

\vspace{10pt}

\noindent\textbf{Funding} 
National Science Foundation (ECCS-1933556, DMR-1747426, RAISE TAQS program); Lawrence Berkeley National Laboratory (Research Subcontract no. 7571809 Prime Award: DE-AC02-05CH11231); National Defense Science and Engineering Graduate Fellowship Program; European Union’s Horizon 2020 research and innovation program (grant agreement No.896401); Air Force Research Laboratory (FA8750-20-2-1007). 

\vspace{10pt}

\noindent\textbf{Acknowledgment}
The U.S. Government is authorized to reproduce and distribute reprints for Governmental purposes notwithstanding any copyright notation thereon.
The views and conclusions contained herein are those of the authors and should not be interpreted as necessarily representing the official policies or endorsements, either expressed or implied, of the Air Force Research Laboratory (AFRL), or the U.S. Government.
This work utilized the MIT Characterization.nano Facility, which provided access to the Raith VELION FIB-SEM system used in this work (Award No. DMR-2117609). We acknowledge the CSEM cleanroom staff for manufacturing the LNOI PIC.

\vspace{10pt}

\noindent\textbf{Disclosures}
Some of the results disclosed in this work were presented at the 2023 Conference on Lasers and Electro Optics (CLEO)~\cite{Errando-Herranz:23}.

\bibliography{MEMSsnspd}

\begin{thebibliography}{37}%
\makeatletter
\providecommand \@ifxundefined [1]{%
 \@ifx{#1\undefined}
}%
\providecommand \@ifnum [1]{%
 \ifnum #1\expandafter \@firstoftwo
 \else \expandafter \@secondoftwo
 \fi
}%
\providecommand \@ifx [1]{%
 \ifx #1\expandafter \@firstoftwo
 \else \expandafter \@secondoftwo
 \fi
}%
\providecommand \natexlab [1]{#1}%
\providecommand \enquote  [1]{``#1''}%
\providecommand \bibnamefont  [1]{#1}%
\providecommand \bibfnamefont [1]{#1}%
\providecommand \citenamefont [1]{#1}%
\providecommand \href@noop [0]{\@secondoftwo}%
\providecommand \href [0]{\begingroup \@sanitize@url \@href}%
\providecommand \@href[1]{\@@startlink{#1}\@@href}%
\providecommand \@@href[1]{\endgroup#1\@@endlink}%
\providecommand \@sanitize@url [0]{\catcode `\\12\catcode `\$12\catcode
  `\&12\catcode `\#12\catcode `\^12\catcode `\_12\catcode `\%12\relax}%
\providecommand \@@startlink[1]{}%
\providecommand \@@endlink[0]{}%
\providecommand \url  [0]{\begingroup\@sanitize@url \@url }%
\providecommand \@url [1]{\endgroup\@href {#1}{\urlprefix }}%
\providecommand \urlprefix  [0]{URL }%
\providecommand \Eprint [0]{\href }%
\providecommand \doibase [0]{http://dx.doi.org/}%
\providecommand \selectlanguage [0]{\@gobble}%
\providecommand \bibinfo  [0]{\@secondoftwo}%
\providecommand \bibfield  [0]{\@secondoftwo}%
\providecommand \translation [1]{[#1]}%
\providecommand \BibitemOpen [0]{}%
\providecommand \bibitemStop [0]{}%
\providecommand \bibitemNoStop [0]{.\EOS\space}%
\providecommand \EOS [0]{\spacefactor3000\relax}%
\providecommand \BibitemShut  [1]{\csname bibitem#1\endcsname}%
\let\auto@bib@innerbib\@empty
\bibitem [{\citenamefont {Ladd}\ \emph {et~al.}(2010)\citenamefont {Ladd},
  \citenamefont {Jelezko}, \citenamefont {Laflamme}, \citenamefont {Nakamura},
  \citenamefont {Monroe},\ and\ \citenamefont {O'Brien}}]{Ladd:10}%
  \BibitemOpen
  \bibfield  {author} {\bibinfo {author} {\bibfnamefont {T.~D.}\ \bibnamefont
  {Ladd}}, \bibinfo {author} {\bibfnamefont {F.}~\bibnamefont {Jelezko}},
  \bibinfo {author} {\bibfnamefont {R.}~\bibnamefont {Laflamme}}, \bibinfo
  {author} {\bibfnamefont {Y.}~\bibnamefont {Nakamura}}, \bibinfo {author}
  {\bibfnamefont {C.}~\bibnamefont {Monroe}}, \ and\ \bibinfo {author}
  {\bibfnamefont {J.~L.}\ \bibnamefont {O'Brien}},\ }\href {\doibase
  10.1038/nature08812} {\bibfield  {journal} {\bibinfo  {journal} {Nature}\
  }\textbf {\bibinfo {volume} {464}},\ \bibinfo {pages} {45} (\bibinfo {year}
  {2010})}\BibitemShut {NoStop}%
\bibitem [{\citenamefont {Gisin}\ and\ \citenamefont {Thew}(2007)}]{Gisin:07}%
  \BibitemOpen
  \bibfield  {author} {\bibinfo {author} {\bibfnamefont {N.}~\bibnamefont
  {Gisin}}\ and\ \bibinfo {author} {\bibfnamefont {R.}~\bibnamefont {Thew}},\
  }\href {\doibase 10.1038/nphoton.2007.22} {\bibfield  {journal} {\bibinfo
  {journal} {Nat. Photon.}\ }\textbf {\bibinfo {volume} {1}},\ \bibinfo {pages}
  {165} (\bibinfo {year} {2007})}\BibitemShut {NoStop}%
\bibitem [{\citenamefont {{Aspuru-Guzik}}\ and\ \citenamefont
  {Walther}(2012)}]{Aspuru-guzik:12}%
  \BibitemOpen
  \bibfield  {author} {\bibinfo {author} {\bibfnamefont {A.}~\bibnamefont
  {{Aspuru-Guzik}}}\ and\ \bibinfo {author} {\bibfnamefont {P.}~\bibnamefont
  {Walther}},\ }\href {\doibase 10.1038/nphys2253} {\bibfield  {journal}
  {\bibinfo  {journal} {Nat. Phys.}\ }\textbf {\bibinfo {volume} {8}},\
  \bibinfo {pages} {285} (\bibinfo {year} {2012})}\BibitemShut {NoStop}%
\bibitem [{\citenamefont {O'Brien}\ \emph {et~al.}(2009)\citenamefont
  {O'Brien}, \citenamefont {Furusawa},\ and\ \citenamefont {Vu{\v
  c}kovi{\'c}}}]{Obrien:09}%
  \BibitemOpen
  \bibfield  {author} {\bibinfo {author} {\bibfnamefont {J.~L.}\ \bibnamefont
  {O'Brien}}, \bibinfo {author} {\bibfnamefont {A.}~\bibnamefont {Furusawa}}, \
  and\ \bibinfo {author} {\bibfnamefont {J.}~\bibnamefont {Vu{\v
  c}kovi{\'c}}},\ }\href {\doibase 10.1038/nphoton.2009.229} {\bibfield
  {journal} {\bibinfo  {journal} {Nat. Photon.}\ }\textbf {\bibinfo {volume}
  {3}},\ \bibinfo {pages} {687} (\bibinfo {year} {2009})}\BibitemShut {NoStop}%
\bibitem [{\citenamefont {Ferrari}\ \emph {et~al.}(2018)\citenamefont
  {Ferrari}, \citenamefont {Schuck},\ and\ \citenamefont
  {Pernice}}]{Ferrari:18}%
  \BibitemOpen
  \bibfield  {author} {\bibinfo {author} {\bibfnamefont {S.}~\bibnamefont
  {Ferrari}}, \bibinfo {author} {\bibfnamefont {C.}~\bibnamefont {Schuck}}, \
  and\ \bibinfo {author} {\bibfnamefont {W.}~\bibnamefont {Pernice}},\ }\href
  {\doibase 10.1515/nanoph-2018-0059} {\bibfield  {journal} {\bibinfo
  {journal} {Nanophotonics}\ }\textbf {\bibinfo {volume} {7}},\ \bibinfo
  {pages} {1725} (\bibinfo {year} {2018})}\BibitemShut {NoStop}%
\bibitem [{\citenamefont {Wang}\ \emph {et~al.}(2020)\citenamefont {Wang},
  \citenamefont {Sciarrino}, \citenamefont {Laing},\ and\ \citenamefont
  {Thompson}}]{Wang:20}%
  \BibitemOpen
  \bibfield  {author} {\bibinfo {author} {\bibfnamefont {J.}~\bibnamefont
  {Wang}}, \bibinfo {author} {\bibfnamefont {F.}~\bibnamefont {Sciarrino}},
  \bibinfo {author} {\bibfnamefont {A.}~\bibnamefont {Laing}}, \ and\ \bibinfo
  {author} {\bibfnamefont {M.~G.}\ \bibnamefont {Thompson}},\ }\href {\doibase
  10.1038/s41566-019-0532-1} {\bibfield  {journal} {\bibinfo  {journal} {Nat.
  Photon.}\ }\textbf {\bibinfo {volume} {14}},\ \bibinfo {pages} {273}
  (\bibinfo {year} {2020})}\BibitemShut {NoStop}%
\bibitem [{\citenamefont {Reddy}\ \emph {et~al.}(2020)\citenamefont {Reddy},
  \citenamefont {Nerem}, \citenamefont {Nam}, \citenamefont {Mirin},\ and\
  \citenamefont {Verma}}]{Reddy:20}%
  \BibitemOpen
  \bibfield  {author} {\bibinfo {author} {\bibfnamefont {D.~V.}\ \bibnamefont
  {Reddy}}, \bibinfo {author} {\bibfnamefont {R.~R.}\ \bibnamefont {Nerem}},
  \bibinfo {author} {\bibfnamefont {S.~W.}\ \bibnamefont {Nam}}, \bibinfo
  {author} {\bibfnamefont {R.~P.}\ \bibnamefont {Mirin}}, \ and\ \bibinfo
  {author} {\bibfnamefont {V.~B.}\ \bibnamefont {Verma}},\ }\href {\doibase
  10.1364/OPTICA.400751} {\bibfield  {journal} {\bibinfo  {journal} {Optica}\
  }\textbf {\bibinfo {volume} {7}},\ \bibinfo {pages} {1649} (\bibinfo {year}
  {2020})}\BibitemShut {NoStop}%
\bibitem [{\citenamefont {Colangelo}\ \emph {et~al.}(2022)\citenamefont
  {Colangelo}, \citenamefont {Walter}, \citenamefont {Korzh}, \citenamefont
  {Schmidt}, \citenamefont {Bumble}, \citenamefont {Lita}, \citenamefont
  {Beyer}, \citenamefont {Allmaras}, \citenamefont {Briggs}, \citenamefont
  {Kozorezov}, \citenamefont {Wollman}, \citenamefont {Shaw},\ and\
  \citenamefont {Berggren}}]{Colangelo:22}%
  \BibitemOpen
  \bibfield  {author} {\bibinfo {author} {\bibfnamefont {M.}~\bibnamefont
  {Colangelo}}, \bibinfo {author} {\bibfnamefont {A.~B.}\ \bibnamefont
  {Walter}}, \bibinfo {author} {\bibfnamefont {B.~A.}\ \bibnamefont {Korzh}},
  \bibinfo {author} {\bibfnamefont {E.}~\bibnamefont {Schmidt}}, \bibinfo
  {author} {\bibfnamefont {B.}~\bibnamefont {Bumble}}, \bibinfo {author}
  {\bibfnamefont {A.~E.}\ \bibnamefont {Lita}}, \bibinfo {author}
  {\bibfnamefont {A.~D.}\ \bibnamefont {Beyer}}, \bibinfo {author}
  {\bibfnamefont {J.~P.}\ \bibnamefont {Allmaras}}, \bibinfo {author}
  {\bibfnamefont {R.~M.}\ \bibnamefont {Briggs}}, \bibinfo {author}
  {\bibfnamefont {A.~G.}\ \bibnamefont {Kozorezov}}, \bibinfo {author}
  {\bibfnamefont {E.~E.}\ \bibnamefont {Wollman}}, \bibinfo {author}
  {\bibfnamefont {M.~D.}\ \bibnamefont {Shaw}}, \ and\ \bibinfo {author}
  {\bibfnamefont {K.~K.}\ \bibnamefont {Berggren}},\ }\href {\doibase
  10.1021/acs.nanolett.1c05012} {\bibfield  {journal} {\bibinfo  {journal}
  {Nano Lett.}\ }\textbf {\bibinfo {volume} {22}},\ \bibinfo {pages} {5667}
  (\bibinfo {year} {2022})}\BibitemShut {NoStop}%
\bibitem [{\citenamefont {Chiles}\ \emph {et~al.}(2022)\citenamefont {Chiles},
  \citenamefont {Charaev}, \citenamefont {Lasenby}, \citenamefont {Baryakhtar},
  \citenamefont {Huang}, \citenamefont {Roshko}, \citenamefont {Burton},
  \citenamefont {Colangelo}, \citenamefont {Van~Tilburg}, \citenamefont
  {Arvanitaki}, \citenamefont {Nam},\ and\ \citenamefont
  {Berggren}}]{Chiles:22}%
  \BibitemOpen
  \bibfield  {author} {\bibinfo {author} {\bibfnamefont {J.}~\bibnamefont
  {Chiles}}, \bibinfo {author} {\bibfnamefont {I.}~\bibnamefont {Charaev}},
  \bibinfo {author} {\bibfnamefont {R.}~\bibnamefont {Lasenby}}, \bibinfo
  {author} {\bibfnamefont {M.}~\bibnamefont {Baryakhtar}}, \bibinfo {author}
  {\bibfnamefont {J.}~\bibnamefont {Huang}}, \bibinfo {author} {\bibfnamefont
  {A.}~\bibnamefont {Roshko}}, \bibinfo {author} {\bibfnamefont
  {G.}~\bibnamefont {Burton}}, \bibinfo {author} {\bibfnamefont
  {M.}~\bibnamefont {Colangelo}}, \bibinfo {author} {\bibfnamefont
  {K.}~\bibnamefont {Van~Tilburg}}, \bibinfo {author} {\bibfnamefont
  {A.}~\bibnamefont {Arvanitaki}}, \bibinfo {author} {\bibfnamefont {S.~W.}\
  \bibnamefont {Nam}}, \ and\ \bibinfo {author} {\bibfnamefont {K.~K.}\
  \bibnamefont {Berggren}},\ }\href {\doibase 10.1103/PhysRevLett.128.231802}
  {\bibfield  {journal} {\bibinfo  {journal} {Phys. Rev. Lett.}\ }\textbf
  {\bibinfo {volume} {128}},\ \bibinfo {pages} {231802} (\bibinfo {year}
  {2022})}\BibitemShut {NoStop}%
\bibitem [{\citenamefont {M{\"u}nzberg}\ \emph {et~al.}(2018)\citenamefont
  {M{\"u}nzberg}, \citenamefont {Vetter}, \citenamefont {Beutel}, \citenamefont
  {Hartmann}, \citenamefont {Ferrari}, \citenamefont {Pernice},\ and\
  \citenamefont {Rockstuhl}}]{Munzberg:18}%
  \BibitemOpen
  \bibfield  {author} {\bibinfo {author} {\bibfnamefont {J.}~\bibnamefont
  {M{\"u}nzberg}}, \bibinfo {author} {\bibfnamefont {A.}~\bibnamefont
  {Vetter}}, \bibinfo {author} {\bibfnamefont {F.}~\bibnamefont {Beutel}},
  \bibinfo {author} {\bibfnamefont {W.}~\bibnamefont {Hartmann}}, \bibinfo
  {author} {\bibfnamefont {S.}~\bibnamefont {Ferrari}}, \bibinfo {author}
  {\bibfnamefont {W.~H.~P.}\ \bibnamefont {Pernice}}, \ and\ \bibinfo {author}
  {\bibfnamefont {C.}~\bibnamefont {Rockstuhl}},\ }\href {\doibase
  10.1364/OPTICA.5.000658} {\bibfield  {journal} {\bibinfo  {journal} {Optica}\
  }\textbf {\bibinfo {volume} {5}},\ \bibinfo {pages} {658} (\bibinfo {year}
  {2018})}\BibitemShut {NoStop}%
\bibitem [{\citenamefont {Gol'tsman}\ \emph {et~al.}(2001)\citenamefont
  {Gol'tsman}, \citenamefont {Okunev}, \citenamefont {Chulkova}, \citenamefont
  {Lipatov}, \citenamefont {Semenov}, \citenamefont {Smirnov}, \citenamefont
  {Voronov}, \citenamefont {Dzardanov}, \citenamefont {Williams},\ and\
  \citenamefont {Sobolewski}}]{Goltsman:01}%
  \BibitemOpen
  \bibfield  {author} {\bibinfo {author} {\bibfnamefont {G.~N.}\ \bibnamefont
  {Gol'tsman}}, \bibinfo {author} {\bibfnamefont {O.}~\bibnamefont {Okunev}},
  \bibinfo {author} {\bibfnamefont {G.}~\bibnamefont {Chulkova}}, \bibinfo
  {author} {\bibfnamefont {A.}~\bibnamefont {Lipatov}}, \bibinfo {author}
  {\bibfnamefont {A.}~\bibnamefont {Semenov}}, \bibinfo {author} {\bibfnamefont
  {K.}~\bibnamefont {Smirnov}}, \bibinfo {author} {\bibfnamefont
  {B.}~\bibnamefont {Voronov}}, \bibinfo {author} {\bibfnamefont
  {A.}~\bibnamefont {Dzardanov}}, \bibinfo {author} {\bibfnamefont
  {C.}~\bibnamefont {Williams}}, \ and\ \bibinfo {author} {\bibfnamefont
  {R.}~\bibnamefont {Sobolewski}},\ }\href {\doibase 10.1063/1.1388868}
  {\bibfield  {journal} {\bibinfo  {journal} {Appl. Phys. Lett.}\ }\textbf
  {\bibinfo {volume} {79}},\ \bibinfo {pages} {705} (\bibinfo {year}
  {2001})}\BibitemShut {NoStop}%
\bibitem [{\citenamefont {Gyger}\ \emph {et~al.}(2021)\citenamefont {Gyger},
  \citenamefont {Zichi}, \citenamefont {Schweickert}, \citenamefont {Elshaari},
  \citenamefont {Steinhauer}, \citenamefont {{Covre da Silva}}, \citenamefont
  {Rastelli}, \citenamefont {Zwiller}, \citenamefont {J{\"o}ns},\ and\
  \citenamefont {{Errando-Herranz}}}]{Gyger:21}%
  \BibitemOpen
  \bibfield  {author} {\bibinfo {author} {\bibfnamefont {S.}~\bibnamefont
  {Gyger}}, \bibinfo {author} {\bibfnamefont {J.}~\bibnamefont {Zichi}},
  \bibinfo {author} {\bibfnamefont {L.}~\bibnamefont {Schweickert}}, \bibinfo
  {author} {\bibfnamefont {A.~W.}\ \bibnamefont {Elshaari}}, \bibinfo {author}
  {\bibfnamefont {S.}~\bibnamefont {Steinhauer}}, \bibinfo {author}
  {\bibfnamefont {S.~F.}\ \bibnamefont {{Covre da Silva}}}, \bibinfo {author}
  {\bibfnamefont {A.}~\bibnamefont {Rastelli}}, \bibinfo {author}
  {\bibfnamefont {V.}~\bibnamefont {Zwiller}}, \bibinfo {author} {\bibfnamefont
  {K.~D.}\ \bibnamefont {J{\"o}ns}}, \ and\ \bibinfo {author} {\bibfnamefont
  {C.}~\bibnamefont {{Errando-Herranz}}},\ }\href {\doibase
  10.1038/s41467-021-21624-3} {\bibfield  {journal} {\bibinfo  {journal} {Nat.
  Commun.}\ }\textbf {\bibinfo {volume} {12}},\ \bibinfo {pages} {1408}
  (\bibinfo {year} {2021})}\BibitemShut {NoStop}%
\bibitem [{\citenamefont {Najafi}\ \emph
  {et~al.}(2015{\natexlab{a}})\citenamefont {Najafi}, \citenamefont {Dane},
  \citenamefont {Bellei}, \citenamefont {Zhao}, \citenamefont {Sunter},
  \citenamefont {McCaughan},\ and\ \citenamefont {Berggren}}]{Najafi:15ii}%
  \BibitemOpen
  \bibfield  {author} {\bibinfo {author} {\bibfnamefont {F.}~\bibnamefont
  {Najafi}}, \bibinfo {author} {\bibfnamefont {A.}~\bibnamefont {Dane}},
  \bibinfo {author} {\bibfnamefont {F.}~\bibnamefont {Bellei}}, \bibinfo
  {author} {\bibfnamefont {Q.}~\bibnamefont {Zhao}}, \bibinfo {author}
  {\bibfnamefont {K.~A.}\ \bibnamefont {Sunter}}, \bibinfo {author}
  {\bibfnamefont {A.~N.}\ \bibnamefont {McCaughan}}, \ and\ \bibinfo {author}
  {\bibfnamefont {K.~K.}\ \bibnamefont {Berggren}},\ }\href {\doibase
  10.1109/JSTQE.2014.2372054} {\bibfield  {journal} {\bibinfo  {journal} {IEEE
  J. Sel. Top. Quantum Electron.}\ }\textbf {\bibinfo {volume} {21}},\ \bibinfo
  {pages} {1} (\bibinfo {year} {2015}{\natexlab{a}})}\BibitemShut {NoStop}%
\bibitem [{\citenamefont {Cheng}\ \emph {et~al.}(2019)\citenamefont {Cheng},
  \citenamefont {Wang},\ and\ \citenamefont {Tang}}]{Cheng:19}%
  \BibitemOpen
  \bibfield  {author} {\bibinfo {author} {\bibfnamefont {R.}~\bibnamefont
  {Cheng}}, \bibinfo {author} {\bibfnamefont {S.}~\bibnamefont {Wang}}, \ and\
  \bibinfo {author} {\bibfnamefont {H.~X.}\ \bibnamefont {Tang}},\ }\href
  {\doibase 10.1063/1.5131664} {\bibfield  {journal} {\bibinfo  {journal}
  {Appl. Phys. Lett.}\ }\textbf {\bibinfo {volume} {115}},\ \bibinfo {pages}
  {241101} (\bibinfo {year} {2019})}\BibitemShut {NoStop}%
\bibitem [{\citenamefont {Elshaari}\ \emph {et~al.}(2020)\citenamefont
  {Elshaari}, \citenamefont {Pernice}, \citenamefont {Srinivasan},
  \citenamefont {Benson},\ and\ \citenamefont {Zwiller}}]{Elshaari:20}%
  \BibitemOpen
  \bibfield  {author} {\bibinfo {author} {\bibfnamefont {A.~W.}\ \bibnamefont
  {Elshaari}}, \bibinfo {author} {\bibfnamefont {W.}~\bibnamefont {Pernice}},
  \bibinfo {author} {\bibfnamefont {K.}~\bibnamefont {Srinivasan}}, \bibinfo
  {author} {\bibfnamefont {O.}~\bibnamefont {Benson}}, \ and\ \bibinfo {author}
  {\bibfnamefont {V.}~\bibnamefont {Zwiller}},\ }\href {\doibase
  10.1038/s41566-020-0609-x} {\bibfield  {journal} {\bibinfo  {journal} {Nat.
  Photon.}\ ,\ \bibinfo {pages} {1}} (\bibinfo {year} {2020})}\BibitemShut
  {NoStop}%
\bibitem [{\citenamefont {Kim}\ \emph {et~al.}(2020)\citenamefont {Kim},
  \citenamefont {Aghaeimeibodi}, \citenamefont {Carolan}, \citenamefont
  {Englund},\ and\ \citenamefont {Waks}}]{Kim:20}%
  \BibitemOpen
  \bibfield  {author} {\bibinfo {author} {\bibfnamefont {J.-H.}\ \bibnamefont
  {Kim}}, \bibinfo {author} {\bibfnamefont {S.}~\bibnamefont {Aghaeimeibodi}},
  \bibinfo {author} {\bibfnamefont {J.}~\bibnamefont {Carolan}}, \bibinfo
  {author} {\bibfnamefont {D.}~\bibnamefont {Englund}}, \ and\ \bibinfo
  {author} {\bibfnamefont {E.}~\bibnamefont {Waks}},\ }\href {\doibase
  10.1364/OPTICA.384118} {\bibfield  {journal} {\bibinfo  {journal} {Optica}\
  }\textbf {\bibinfo {volume} {7}},\ \bibinfo {pages} {291} (\bibinfo {year}
  {2020})}\BibitemShut {NoStop}%
\bibitem [{\citenamefont {Najafi}\ \emph
  {et~al.}(2015{\natexlab{b}})\citenamefont {Najafi}, \citenamefont {Mower},
  \citenamefont {Harris}, \citenamefont {Bellei}, \citenamefont {Dane},
  \citenamefont {Lee}, \citenamefont {Hu}, \citenamefont {Kharel},
  \citenamefont {Marsili}, \citenamefont {Assefa}, \citenamefont {Berggren},\
  and\ \citenamefont {Englund}}]{Najafi:15}%
  \BibitemOpen
  \bibfield  {author} {\bibinfo {author} {\bibfnamefont {F.}~\bibnamefont
  {Najafi}}, \bibinfo {author} {\bibfnamefont {J.}~\bibnamefont {Mower}},
  \bibinfo {author} {\bibfnamefont {N.~C.}\ \bibnamefont {Harris}}, \bibinfo
  {author} {\bibfnamefont {F.}~\bibnamefont {Bellei}}, \bibinfo {author}
  {\bibfnamefont {A.}~\bibnamefont {Dane}}, \bibinfo {author} {\bibfnamefont
  {C.}~\bibnamefont {Lee}}, \bibinfo {author} {\bibfnamefont {X.}~\bibnamefont
  {Hu}}, \bibinfo {author} {\bibfnamefont {P.}~\bibnamefont {Kharel}}, \bibinfo
  {author} {\bibfnamefont {F.}~\bibnamefont {Marsili}}, \bibinfo {author}
  {\bibfnamefont {S.}~\bibnamefont {Assefa}}, \bibinfo {author} {\bibfnamefont
  {K.~K.}\ \bibnamefont {Berggren}}, \ and\ \bibinfo {author} {\bibfnamefont
  {D.}~\bibnamefont {Englund}},\ }\href {\doibase 10.1038/ncomms6873}
  {\bibfield  {journal} {\bibinfo  {journal} {Nat. Commun.}\ }\textbf {\bibinfo
  {volume} {6}},\ \bibinfo {pages} {5873} (\bibinfo {year}
  {2015}{\natexlab{b}})}\BibitemShut {NoStop}%
\bibitem [{\citenamefont {Sayem}\ \emph {et~al.}(2020)\citenamefont {Sayem},
  \citenamefont {Cheng}, \citenamefont {Wang},\ and\ \citenamefont
  {Tang}}]{Sayem:20}%
  \BibitemOpen
  \bibfield  {author} {\bibinfo {author} {\bibfnamefont {A.~A.}\ \bibnamefont
  {Sayem}}, \bibinfo {author} {\bibfnamefont {R.}~\bibnamefont {Cheng}},
  \bibinfo {author} {\bibfnamefont {S.}~\bibnamefont {Wang}}, \ and\ \bibinfo
  {author} {\bibfnamefont {H.~X.}\ \bibnamefont {Tang}},\ }\href {\doibase
  10.1063/1.5142852} {\bibfield  {journal} {\bibinfo  {journal} {Appl. Phys.
  Lett.}\ }\textbf {\bibinfo {volume} {116}},\ \bibinfo {pages} {151102}
  (\bibinfo {year} {2020})}\BibitemShut {NoStop}%
\bibitem [{\citenamefont {Lomonte}\ \emph {et~al.}(2021)\citenamefont
  {Lomonte}, \citenamefont {Wolff}, \citenamefont {Beutel}, \citenamefont
  {Ferrari}, \citenamefont {Schuck}, \citenamefont {Pernice},\ and\
  \citenamefont {Lenzini}}]{Lomonte:21}%
  \BibitemOpen
  \bibfield  {author} {\bibinfo {author} {\bibfnamefont {E.}~\bibnamefont
  {Lomonte}}, \bibinfo {author} {\bibfnamefont {M.~A.}\ \bibnamefont {Wolff}},
  \bibinfo {author} {\bibfnamefont {F.}~\bibnamefont {Beutel}}, \bibinfo
  {author} {\bibfnamefont {S.}~\bibnamefont {Ferrari}}, \bibinfo {author}
  {\bibfnamefont {C.}~\bibnamefont {Schuck}}, \bibinfo {author} {\bibfnamefont
  {W.~H.~P.}\ \bibnamefont {Pernice}}, \ and\ \bibinfo {author} {\bibfnamefont
  {F.}~\bibnamefont {Lenzini}},\ }\href {\doibase 10.1038/s41467-021-27205-8}
  {\bibfield  {journal} {\bibinfo  {journal} {Nat. Commun.}\ }\textbf {\bibinfo
  {volume} {12}},\ \bibinfo {pages} {6847} (\bibinfo {year}
  {2021})}\BibitemShut {NoStop}%
\bibitem [{\citenamefont {Colangelo}\ \emph {et~al.}(2024)\citenamefont
  {Colangelo}, \citenamefont {Zhu}, \citenamefont {Shao}, \citenamefont
  {Holzgrafe}, \citenamefont {Batson}, \citenamefont {Desiatov}, \citenamefont
  {Medeiros}, \citenamefont {Yeung}, \citenamefont {Loncar},\ and\
  \citenamefont {Berggren}}]{Colangelo:24}%
  \BibitemOpen
  \bibfield  {author} {\bibinfo {author} {\bibfnamefont {M.}~\bibnamefont
  {Colangelo}}, \bibinfo {author} {\bibfnamefont {D.}~\bibnamefont {Zhu}},
  \bibinfo {author} {\bibfnamefont {L.}~\bibnamefont {Shao}}, \bibinfo {author}
  {\bibfnamefont {J.}~\bibnamefont {Holzgrafe}}, \bibinfo {author}
  {\bibfnamefont {E.~K.}\ \bibnamefont {Batson}}, \bibinfo {author}
  {\bibfnamefont {B.}~\bibnamefont {Desiatov}}, \bibinfo {author}
  {\bibfnamefont {O.}~\bibnamefont {Medeiros}}, \bibinfo {author}
  {\bibfnamefont {M.}~\bibnamefont {Yeung}}, \bibinfo {author} {\bibfnamefont
  {M.}~\bibnamefont {Loncar}}, \ and\ \bibinfo {author} {\bibfnamefont {K.~K.}\
  \bibnamefont {Berggren}},\ }\href {\doibase 10.1021/acsphotonics.3c01628}
  {\bibfield  {journal} {\bibinfo  {journal} {ACS Photonics}\ }\textbf
  {\bibinfo {volume} {11}},\ \bibinfo {pages} {356} (\bibinfo {year}
  {2024})}\BibitemShut {NoStop}%
\bibitem [{\citenamefont {Hochberg}\ and\ \citenamefont
  {Baehr-Jones}(2010)}]{Hochberg:10}%
  \BibitemOpen
  \bibfield  {author} {\bibinfo {author} {\bibfnamefont {M.}~\bibnamefont
  {Hochberg}}\ and\ \bibinfo {author} {\bibfnamefont {T.}~\bibnamefont
  {Baehr-Jones}},\ }\href {\doibase 10.1038/nphoton.2010.172} {\bibfield
  {journal} {\bibinfo  {journal} {Nat. Photon.}\ }\textbf {\bibinfo {volume}
  {4}},\ \bibinfo {pages} {492} (\bibinfo {year} {2010})}\BibitemShut {NoStop}%
\bibitem [{\citenamefont {Fahrenkopf}\ \emph {et~al.}(2019)\citenamefont
  {Fahrenkopf}, \citenamefont {McDonough}, \citenamefont {Leake}, \citenamefont
  {Su}, \citenamefont {Timurdogan},\ and\ \citenamefont
  {Coolbaugh}}]{Fahrenkopf:19}%
  \BibitemOpen
  \bibfield  {author} {\bibinfo {author} {\bibfnamefont {N.~M.}\ \bibnamefont
  {Fahrenkopf}}, \bibinfo {author} {\bibfnamefont {C.}~\bibnamefont
  {McDonough}}, \bibinfo {author} {\bibfnamefont {G.~L.}\ \bibnamefont
  {Leake}}, \bibinfo {author} {\bibfnamefont {Z.}~\bibnamefont {Su}}, \bibinfo
  {author} {\bibfnamefont {E.}~\bibnamefont {Timurdogan}}, \ and\ \bibinfo
  {author} {\bibfnamefont {D.~D.}\ \bibnamefont {Coolbaugh}},\ }\href {\doibase
  10.1109/JSTQE.2019.2935698} {\bibfield  {journal} {\bibinfo  {journal} {IEEE
  J. Sel. Top. Quantum Electron.}\ }\textbf {\bibinfo {volume} {25}},\ \bibinfo
  {pages} {1} (\bibinfo {year} {2019})}\BibitemShut {NoStop}%
\bibitem [{\citenamefont {Alexander}\ \emph {et~al.}(2024)\citenamefont
  {Alexander}, \citenamefont {Bahgat}, \citenamefont {Benyamini}, \citenamefont
  {Black}, \citenamefont {Bonneau}, \citenamefont {Burgos}, \citenamefont
  {Burridge}, \citenamefont {Campbell}, \citenamefont {Catalano}, \citenamefont
  {Ceballos}, \citenamefont {Chang}, \citenamefont {Chung}, \citenamefont
  {Danesh}, \citenamefont {Dauer}, \citenamefont {Davis}, \citenamefont
  {Dudley}, \citenamefont {Er-Xuan}, \citenamefont {Fargas}, \citenamefont
  {Farsi}, \citenamefont {Fenrich}, \citenamefont {Frazer}, \citenamefont
  {Fukami}, \citenamefont {Ganesan}, \citenamefont {Gibson}, \citenamefont
  {Gimeno-Segovia}, \citenamefont {Goeldi}, \citenamefont {Goley},
  \citenamefont {Haislmaier}, \citenamefont {Halimi}, \citenamefont {Hansen},
  \citenamefont {Hardy}, \citenamefont {Horng}, \citenamefont {House},
  \citenamefont {Hu}, \citenamefont {Jadidi}, \citenamefont {Johansson},
  \citenamefont {Jones}, \citenamefont {Kamineni}, \citenamefont {Kelez},
  \citenamefont {Koustuban}, \citenamefont {Kovall}, \citenamefont {Krogen},
  \citenamefont {Kumar}, \citenamefont {Liang}, \citenamefont {LiCausi},
  \citenamefont {Llewellyn}, \citenamefont {Lokovic}, \citenamefont {Lovelady},
  \citenamefont {Manfrinato}, \citenamefont {Melnichuk}, \citenamefont {Souza},
  \citenamefont {Mendoza}, \citenamefont {Moores}, \citenamefont {Mukherjee},
  \citenamefont {Munns}, \citenamefont {Musalem}, \citenamefont {Najafi},
  \citenamefont {O'Brien}, \citenamefont {Ortmann}, \citenamefont {Pai},
  \citenamefont {Park}, \citenamefont {Peng}, \citenamefont {Penthorn},
  \citenamefont {Peterson}, \citenamefont {Poush}, \citenamefont {Pryde},
  \citenamefont {Ramprasad}, \citenamefont {Ray}, \citenamefont {Rodriguez},
  \citenamefont {Roxworthy}, \citenamefont {Rudolph}, \citenamefont {Saunders},
  \citenamefont {Shadbolt}, \citenamefont {Shah}, \citenamefont {Shin},
  \citenamefont {Smith}, \citenamefont {Sohn}, \citenamefont {Sohn},
  \citenamefont {Son}, \citenamefont {Sparrow}, \citenamefont {Staffaroni},
  \citenamefont {Stavrakas}, \citenamefont {Sukumaran}, \citenamefont
  {Tamborini}, \citenamefont {Thompson}, \citenamefont {Tran}, \citenamefont
  {Triplet}, \citenamefont {Tung}, \citenamefont {Vert}, \citenamefont
  {Vidrighin}, \citenamefont {Vorobeichik}, \citenamefont {Weigel},
  \citenamefont {Wingert}, \citenamefont {Wooding},\ and\ \citenamefont
  {Zhou}}]{Alexander:24}%
  \BibitemOpen
  \bibfield  {author} {\bibinfo {author} {\bibfnamefont {K.}~\bibnamefont
  {Alexander}}, \bibinfo {author} {\bibfnamefont {A.}~\bibnamefont {Bahgat}},
  \bibinfo {author} {\bibfnamefont {A.}~\bibnamefont {Benyamini}}, \bibinfo
  {author} {\bibfnamefont {D.}~\bibnamefont {Black}}, \bibinfo {author}
  {\bibfnamefont {D.}~\bibnamefont {Bonneau}}, \bibinfo {author} {\bibfnamefont
  {S.}~\bibnamefont {Burgos}}, \bibinfo {author} {\bibfnamefont
  {B.}~\bibnamefont {Burridge}}, \bibinfo {author} {\bibfnamefont
  {G.}~\bibnamefont {Campbell}}, \bibinfo {author} {\bibfnamefont
  {G.}~\bibnamefont {Catalano}}, \bibinfo {author} {\bibfnamefont
  {A.}~\bibnamefont {Ceballos}}, \bibinfo {author} {\bibfnamefont {C.-M.}\
  \bibnamefont {Chang}}, \bibinfo {author} {\bibfnamefont {C.}~\bibnamefont
  {Chung}}, \bibinfo {author} {\bibfnamefont {F.}~\bibnamefont {Danesh}},
  \bibinfo {author} {\bibfnamefont {T.}~\bibnamefont {Dauer}}, \bibinfo
  {author} {\bibfnamefont {M.}~\bibnamefont {Davis}}, \bibinfo {author}
  {\bibfnamefont {E.}~\bibnamefont {Dudley}}, \bibinfo {author} {\bibfnamefont
  {P.}~\bibnamefont {Er-Xuan}}, \bibinfo {author} {\bibfnamefont
  {J.}~\bibnamefont {Fargas}}, \bibinfo {author} {\bibfnamefont
  {A.}~\bibnamefont {Farsi}}, \bibinfo {author} {\bibfnamefont
  {C.}~\bibnamefont {Fenrich}}, \bibinfo {author} {\bibfnamefont
  {J.}~\bibnamefont {Frazer}}, \bibinfo {author} {\bibfnamefont
  {M.}~\bibnamefont {Fukami}}, \bibinfo {author} {\bibfnamefont
  {Y.}~\bibnamefont {Ganesan}}, \bibinfo {author} {\bibfnamefont
  {G.}~\bibnamefont {Gibson}}, \bibinfo {author} {\bibfnamefont
  {M.}~\bibnamefont {Gimeno-Segovia}}, \bibinfo {author} {\bibfnamefont
  {S.}~\bibnamefont {Goeldi}}, \bibinfo {author} {\bibfnamefont
  {P.}~\bibnamefont {Goley}}, \bibinfo {author} {\bibfnamefont
  {R.}~\bibnamefont {Haislmaier}}, \bibinfo {author} {\bibfnamefont
  {S.}~\bibnamefont {Halimi}}, \bibinfo {author} {\bibfnamefont
  {P.}~\bibnamefont {Hansen}}, \bibinfo {author} {\bibfnamefont
  {S.}~\bibnamefont {Hardy}}, \bibinfo {author} {\bibfnamefont
  {J.}~\bibnamefont {Horng}}, \bibinfo {author} {\bibfnamefont
  {M.}~\bibnamefont {House}}, \bibinfo {author} {\bibfnamefont
  {H.}~\bibnamefont {Hu}}, \bibinfo {author} {\bibfnamefont {M.}~\bibnamefont
  {Jadidi}}, \bibinfo {author} {\bibfnamefont {H.}~\bibnamefont {Johansson}},
  \bibinfo {author} {\bibfnamefont {T.}~\bibnamefont {Jones}}, \bibinfo
  {author} {\bibfnamefont {V.}~\bibnamefont {Kamineni}}, \bibinfo {author}
  {\bibfnamefont {N.}~\bibnamefont {Kelez}}, \bibinfo {author} {\bibfnamefont
  {R.}~\bibnamefont {Koustuban}}, \bibinfo {author} {\bibfnamefont
  {G.}~\bibnamefont {Kovall}}, \bibinfo {author} {\bibfnamefont
  {P.}~\bibnamefont {Krogen}}, \bibinfo {author} {\bibfnamefont
  {N.}~\bibnamefont {Kumar}}, \bibinfo {author} {\bibfnamefont
  {Y.}~\bibnamefont {Liang}}, \bibinfo {author} {\bibfnamefont
  {N.}~\bibnamefont {LiCausi}}, \bibinfo {author} {\bibfnamefont
  {D.}~\bibnamefont {Llewellyn}}, \bibinfo {author} {\bibfnamefont
  {K.}~\bibnamefont {Lokovic}}, \bibinfo {author} {\bibfnamefont
  {M.}~\bibnamefont {Lovelady}}, \bibinfo {author} {\bibfnamefont
  {V.}~\bibnamefont {Manfrinato}}, \bibinfo {author} {\bibfnamefont
  {A.}~\bibnamefont {Melnichuk}}, \bibinfo {author} {\bibfnamefont
  {M.}~\bibnamefont {Souza}}, \bibinfo {author} {\bibfnamefont
  {G.}~\bibnamefont {Mendoza}}, \bibinfo {author} {\bibfnamefont
  {B.}~\bibnamefont {Moores}}, \bibinfo {author} {\bibfnamefont
  {S.}~\bibnamefont {Mukherjee}}, \bibinfo {author} {\bibfnamefont
  {J.}~\bibnamefont {Munns}}, \bibinfo {author} {\bibfnamefont {F.-X.}\
  \bibnamefont {Musalem}}, \bibinfo {author} {\bibfnamefont {F.}~\bibnamefont
  {Najafi}}, \bibinfo {author} {\bibfnamefont {J.~L.}\ \bibnamefont {O'Brien}},
  \bibinfo {author} {\bibfnamefont {J.~E.}\ \bibnamefont {Ortmann}}, \bibinfo
  {author} {\bibfnamefont {S.}~\bibnamefont {Pai}}, \bibinfo {author}
  {\bibfnamefont {B.}~\bibnamefont {Park}}, \bibinfo {author} {\bibfnamefont
  {H.-T.}\ \bibnamefont {Peng}}, \bibinfo {author} {\bibfnamefont
  {N.}~\bibnamefont {Penthorn}}, \bibinfo {author} {\bibfnamefont
  {B.}~\bibnamefont {Peterson}}, \bibinfo {author} {\bibfnamefont
  {M.}~\bibnamefont {Poush}}, \bibinfo {author} {\bibfnamefont {G.~J.}\
  \bibnamefont {Pryde}}, \bibinfo {author} {\bibfnamefont {T.}~\bibnamefont
  {Ramprasad}}, \bibinfo {author} {\bibfnamefont {G.}~\bibnamefont {Ray}},
  \bibinfo {author} {\bibfnamefont {A.}~\bibnamefont {Rodriguez}}, \bibinfo
  {author} {\bibfnamefont {B.}~\bibnamefont {Roxworthy}}, \bibinfo {author}
  {\bibfnamefont {T.}~\bibnamefont {Rudolph}}, \bibinfo {author} {\bibfnamefont
  {D.~J.}\ \bibnamefont {Saunders}}, \bibinfo {author} {\bibfnamefont
  {P.}~\bibnamefont {Shadbolt}}, \bibinfo {author} {\bibfnamefont
  {D.}~\bibnamefont {Shah}}, \bibinfo {author} {\bibfnamefont {H.}~\bibnamefont
  {Shin}}, \bibinfo {author} {\bibfnamefont {J.}~\bibnamefont {Smith}},
  \bibinfo {author} {\bibfnamefont {B.}~\bibnamefont {Sohn}}, \bibinfo {author}
  {\bibfnamefont {Y.-I.}\ \bibnamefont {Sohn}}, \bibinfo {author}
  {\bibfnamefont {G.}~\bibnamefont {Son}}, \bibinfo {author} {\bibfnamefont
  {C.}~\bibnamefont {Sparrow}}, \bibinfo {author} {\bibfnamefont
  {M.}~\bibnamefont {Staffaroni}}, \bibinfo {author} {\bibfnamefont
  {C.}~\bibnamefont {Stavrakas}}, \bibinfo {author} {\bibfnamefont
  {V.}~\bibnamefont {Sukumaran}}, \bibinfo {author} {\bibfnamefont
  {D.}~\bibnamefont {Tamborini}}, \bibinfo {author} {\bibfnamefont {M.~G.}\
  \bibnamefont {Thompson}}, \bibinfo {author} {\bibfnamefont {K.}~\bibnamefont
  {Tran}}, \bibinfo {author} {\bibfnamefont {M.}~\bibnamefont {Triplet}},
  \bibinfo {author} {\bibfnamefont {M.}~\bibnamefont {Tung}}, \bibinfo {author}
  {\bibfnamefont {A.}~\bibnamefont {Vert}}, \bibinfo {author} {\bibfnamefont
  {M.~D.}\ \bibnamefont {Vidrighin}}, \bibinfo {author} {\bibfnamefont
  {I.}~\bibnamefont {Vorobeichik}}, \bibinfo {author} {\bibfnamefont
  {P.}~\bibnamefont {Weigel}}, \bibinfo {author} {\bibfnamefont
  {M.}~\bibnamefont {Wingert}}, \bibinfo {author} {\bibfnamefont
  {J.}~\bibnamefont {Wooding}}, \ and\ \bibinfo {author} {\bibfnamefont
  {X.}~\bibnamefont {Zhou}},\ }\href@noop {} {\enquote {\bibinfo {title} {A
  manufacturable platform for photonic quantum computing},}\ }\bibinfo
  {howpublished} {Preprint at https://arxiv.org/abs/2404.17570} (\bibinfo
  {year} {2024})\BibitemShut {NoStop}%
\bibitem [{\citenamefont {Justice}\ \emph {et~al.}(2012)\citenamefont
  {Justice}, \citenamefont {Bower}, \citenamefont {Meitl}, \citenamefont
  {Mooney}, \citenamefont {Gubbins},\ and\ \citenamefont
  {Corbett}}]{Justice:12}%
  \BibitemOpen
  \bibfield  {author} {\bibinfo {author} {\bibfnamefont {J.}~\bibnamefont
  {Justice}}, \bibinfo {author} {\bibfnamefont {C.}~\bibnamefont {Bower}},
  \bibinfo {author} {\bibfnamefont {M.}~\bibnamefont {Meitl}}, \bibinfo
  {author} {\bibfnamefont {M.~B.}\ \bibnamefont {Mooney}}, \bibinfo {author}
  {\bibfnamefont {M.~A.}\ \bibnamefont {Gubbins}}, \ and\ \bibinfo {author}
  {\bibfnamefont {B.}~\bibnamefont {Corbett}},\ }\href {\doibase
  10.1038/nphoton.2012.204} {\bibfield  {journal} {\bibinfo  {journal} {Nat.
  Photon.}\ }\textbf {\bibinfo {volume} {6}},\ \bibinfo {pages} {610} (\bibinfo
  {year} {2012})}\BibitemShut {NoStop}%
\bibitem [{\citenamefont {Larocque}\ \emph {et~al.}(2024)\citenamefont
  {Larocque}, \citenamefont {Buyukkaya}, \citenamefont {Errando-Herranz},
  \citenamefont {Papon}, \citenamefont {Harper}, \citenamefont {Tao},
  \citenamefont {Carolan}, \citenamefont {Lee}, \citenamefont {Richardson},
  \citenamefont {Leake}, \citenamefont {Coleman}, \citenamefont {Fanto},
  \citenamefont {Waks},\ and\ \citenamefont {Englund}}]{Larocque:24}%
  \BibitemOpen
  \bibfield  {author} {\bibinfo {author} {\bibfnamefont {H.}~\bibnamefont
  {Larocque}}, \bibinfo {author} {\bibfnamefont {M.~A.}\ \bibnamefont
  {Buyukkaya}}, \bibinfo {author} {\bibfnamefont {C.}~\bibnamefont
  {Errando-Herranz}}, \bibinfo {author} {\bibfnamefont {C.}~\bibnamefont
  {Papon}}, \bibinfo {author} {\bibfnamefont {S.}~\bibnamefont {Harper}},
  \bibinfo {author} {\bibfnamefont {M.}~\bibnamefont {Tao}}, \bibinfo {author}
  {\bibfnamefont {J.}~\bibnamefont {Carolan}}, \bibinfo {author} {\bibfnamefont
  {C.-M.}\ \bibnamefont {Lee}}, \bibinfo {author} {\bibfnamefont {C.~J.~K.}\
  \bibnamefont {Richardson}}, \bibinfo {author} {\bibfnamefont {G.~L.}\
  \bibnamefont {Leake}}, \bibinfo {author} {\bibfnamefont {D.~J.}\ \bibnamefont
  {Coleman}}, \bibinfo {author} {\bibfnamefont {M.~L.}\ \bibnamefont {Fanto}},
  \bibinfo {author} {\bibfnamefont {E.}~\bibnamefont {Waks}}, \ and\ \bibinfo
  {author} {\bibfnamefont {D.}~\bibnamefont {Englund}},\ }\href {\doibase
  10.1038/s41467-024-50208-0} {\bibfield  {journal} {\bibinfo  {journal} {Nat.
  Commun.}\ }\textbf {\bibinfo {volume} {15}},\ \bibinfo {pages} {5781}
  (\bibinfo {year} {2024})}\BibitemShut {NoStop}%
\bibitem [{\citenamefont {Tanner}\ \emph {et~al.}(2012)\citenamefont {Tanner},
  \citenamefont {Alvarez}, \citenamefont {Jiang}, \citenamefont {Warburton},
  \citenamefont {Barber},\ and\ \citenamefont {Hadfield}}]{Tanner:12}%
  \BibitemOpen
  \bibfield  {author} {\bibinfo {author} {\bibfnamefont {M.~G.}\ \bibnamefont
  {Tanner}}, \bibinfo {author} {\bibfnamefont {L.~S.~E.}\ \bibnamefont
  {Alvarez}}, \bibinfo {author} {\bibfnamefont {W.}~\bibnamefont {Jiang}},
  \bibinfo {author} {\bibfnamefont {R.~J.}\ \bibnamefont {Warburton}}, \bibinfo
  {author} {\bibfnamefont {Z.~H.}\ \bibnamefont {Barber}}, \ and\ \bibinfo
  {author} {\bibfnamefont {R.~H.}\ \bibnamefont {Hadfield}},\ }\href {\doibase
  10.1088/0957-4484/23/50/505201} {\bibfield  {journal} {\bibinfo  {journal}
  {Nanotechnology}\ }\textbf {\bibinfo {volume} {23}},\ \bibinfo {pages}
  {505201} (\bibinfo {year} {2012})}\BibitemShut {NoStop}%
\bibitem [{\citenamefont {Korzh}\ \emph {et~al.}(2020)\citenamefont {Korzh},
  \citenamefont {Zhao}, \citenamefont {Allmaras}, \citenamefont {Frasca},
  \citenamefont {Autry}, \citenamefont {Bersin}, \citenamefont {Beyer},
  \citenamefont {Briggs}, \citenamefont {Bumble}, \citenamefont {Colangelo},
  \citenamefont {Crouch}, \citenamefont {Dane}, \citenamefont {Gerrits},
  \citenamefont {Lita}, \citenamefont {Marsili}, \citenamefont {Moody},
  \citenamefont {Pe{\~n}a}, \citenamefont {Ramirez}, \citenamefont {Rezac},
  \citenamefont {Sinclair}, \citenamefont {Stevens}, \citenamefont {Velasco},
  \citenamefont {Verma}, \citenamefont {Wollman}, \citenamefont {Xie},
  \citenamefont {Zhu}, \citenamefont {Hale}, \citenamefont {Spiropulu},
  \citenamefont {Silverman}, \citenamefont {Mirin}, \citenamefont {Nam},
  \citenamefont {Kozorezov}, \citenamefont {Shaw},\ and\ \citenamefont
  {Berggren}}]{Korzh:20}%
  \BibitemOpen
  \bibfield  {author} {\bibinfo {author} {\bibfnamefont {B.}~\bibnamefont
  {Korzh}}, \bibinfo {author} {\bibfnamefont {Q.-Y.}\ \bibnamefont {Zhao}},
  \bibinfo {author} {\bibfnamefont {J.~P.}\ \bibnamefont {Allmaras}}, \bibinfo
  {author} {\bibfnamefont {S.}~\bibnamefont {Frasca}}, \bibinfo {author}
  {\bibfnamefont {T.~M.}\ \bibnamefont {Autry}}, \bibinfo {author}
  {\bibfnamefont {E.~A.}\ \bibnamefont {Bersin}}, \bibinfo {author}
  {\bibfnamefont {A.~D.}\ \bibnamefont {Beyer}}, \bibinfo {author}
  {\bibfnamefont {R.~M.}\ \bibnamefont {Briggs}}, \bibinfo {author}
  {\bibfnamefont {B.}~\bibnamefont {Bumble}}, \bibinfo {author} {\bibfnamefont
  {M.}~\bibnamefont {Colangelo}}, \bibinfo {author} {\bibfnamefont {G.~M.}\
  \bibnamefont {Crouch}}, \bibinfo {author} {\bibfnamefont {A.~E.}\
  \bibnamefont {Dane}}, \bibinfo {author} {\bibfnamefont {T.}~\bibnamefont
  {Gerrits}}, \bibinfo {author} {\bibfnamefont {A.~E.}\ \bibnamefont {Lita}},
  \bibinfo {author} {\bibfnamefont {F.}~\bibnamefont {Marsili}}, \bibinfo
  {author} {\bibfnamefont {G.}~\bibnamefont {Moody}}, \bibinfo {author}
  {\bibfnamefont {C.}~\bibnamefont {Pe{\~n}a}}, \bibinfo {author}
  {\bibfnamefont {E.}~\bibnamefont {Ramirez}}, \bibinfo {author} {\bibfnamefont
  {J.~D.}\ \bibnamefont {Rezac}}, \bibinfo {author} {\bibfnamefont
  {N.}~\bibnamefont {Sinclair}}, \bibinfo {author} {\bibfnamefont {M.~J.}\
  \bibnamefont {Stevens}}, \bibinfo {author} {\bibfnamefont {A.~E.}\
  \bibnamefont {Velasco}}, \bibinfo {author} {\bibfnamefont {V.~B.}\
  \bibnamefont {Verma}}, \bibinfo {author} {\bibfnamefont {E.~E.}\ \bibnamefont
  {Wollman}}, \bibinfo {author} {\bibfnamefont {S.}~\bibnamefont {Xie}},
  \bibinfo {author} {\bibfnamefont {D.}~\bibnamefont {Zhu}}, \bibinfo {author}
  {\bibfnamefont {P.~D.}\ \bibnamefont {Hale}}, \bibinfo {author}
  {\bibfnamefont {M.}~\bibnamefont {Spiropulu}}, \bibinfo {author}
  {\bibfnamefont {K.~L.}\ \bibnamefont {Silverman}}, \bibinfo {author}
  {\bibfnamefont {R.~P.}\ \bibnamefont {Mirin}}, \bibinfo {author}
  {\bibfnamefont {S.~W.}\ \bibnamefont {Nam}}, \bibinfo {author} {\bibfnamefont
  {A.~G.}\ \bibnamefont {Kozorezov}}, \bibinfo {author} {\bibfnamefont {M.~D.}\
  \bibnamefont {Shaw}}, \ and\ \bibinfo {author} {\bibfnamefont {K.~K.}\
  \bibnamefont {Berggren}},\ }\href {\doibase 10.1038/s41566-020-0589-x}
  {\bibfield  {journal} {\bibinfo  {journal} {Nat. Photon.}\ }\textbf {\bibinfo
  {volume} {14}},\ \bibinfo {pages} {250} (\bibinfo {year} {2020})}\BibitemShut
  {NoStop}%
\bibitem [{\citenamefont {Pernice}\ \emph {et~al.}(2012)\citenamefont
  {Pernice}, \citenamefont {Schuck}, \citenamefont {Minaeva}, \citenamefont
  {Li}, \citenamefont {Goltsman}, \citenamefont {Sergienko},\ and\
  \citenamefont {Tang}}]{Pernice:12}%
  \BibitemOpen
  \bibfield  {author} {\bibinfo {author} {\bibfnamefont {W.~H.~P.}\
  \bibnamefont {Pernice}}, \bibinfo {author} {\bibfnamefont {C.}~\bibnamefont
  {Schuck}}, \bibinfo {author} {\bibfnamefont {O.}~\bibnamefont {Minaeva}},
  \bibinfo {author} {\bibfnamefont {M.}~\bibnamefont {Li}}, \bibinfo {author}
  {\bibfnamefont {G.~N.}\ \bibnamefont {Goltsman}}, \bibinfo {author}
  {\bibfnamefont {A.~V.}\ \bibnamefont {Sergienko}}, \ and\ \bibinfo {author}
  {\bibfnamefont {H.~X.}\ \bibnamefont {Tang}},\ }\href {\doibase
  10.1038/ncomms2307} {\bibfield  {journal} {\bibinfo  {journal} {Nat.
  Commun.}\ }\textbf {\bibinfo {volume} {3}},\ \bibinfo {pages} {1} (\bibinfo
  {year} {2012})}\BibitemShut {NoStop}%
\bibitem [{\citenamefont {Bandyopadhyay}\ and\ \citenamefont
  {Englund}(2021)}]{Bandyopadhyay:21}%
  \BibitemOpen
  \bibfield  {author} {\bibinfo {author} {\bibfnamefont {S.}~\bibnamefont
  {Bandyopadhyay}}\ and\ \bibinfo {author} {\bibfnamefont {D.}~\bibnamefont
  {Englund}},\ }\href {\doibase 10.48550/arXiv.2110.12851} {\enquote {\bibinfo
  {title} {Alignment-free photonic interconnects},}\ }\bibinfo {howpublished}
  {Preprint at https://arxiv.org/abs/2110.12851} (\bibinfo {year}
  {2021})\BibitemShut {NoStop}%
\bibitem [{\citenamefont {Meitl}\ \emph {et~al.}(2006)\citenamefont {Meitl},
  \citenamefont {Zhu}, \citenamefont {Kumar}, \citenamefont {Lee},
  \citenamefont {Feng}, \citenamefont {Huang}, \citenamefont {Adesida},
  \citenamefont {Nuzzo},\ and\ \citenamefont {Rogers}}]{Meitl:06}%
  \BibitemOpen
  \bibfield  {author} {\bibinfo {author} {\bibfnamefont {M.~A.}\ \bibnamefont
  {Meitl}}, \bibinfo {author} {\bibfnamefont {Z.-T.}\ \bibnamefont {Zhu}},
  \bibinfo {author} {\bibfnamefont {V.}~\bibnamefont {Kumar}}, \bibinfo
  {author} {\bibfnamefont {K.~J.}\ \bibnamefont {Lee}}, \bibinfo {author}
  {\bibfnamefont {X.}~\bibnamefont {Feng}}, \bibinfo {author} {\bibfnamefont
  {Y.~Y.}\ \bibnamefont {Huang}}, \bibinfo {author} {\bibfnamefont
  {I.}~\bibnamefont {Adesida}}, \bibinfo {author} {\bibfnamefont {R.~G.}\
  \bibnamefont {Nuzzo}}, \ and\ \bibinfo {author} {\bibfnamefont {J.~A.}\
  \bibnamefont {Rogers}},\ }\href {\doibase 10.1038/nmat1532} {\bibfield
  {journal} {\bibinfo  {journal} {Nat. Mater.}\ }\textbf {\bibinfo {volume}
  {5}},\ \bibinfo {pages} {33} (\bibinfo {year} {2006})}\BibitemShut {NoStop}%
\bibitem [{\citenamefont {Bower}\ \emph {et~al.}(2018)\citenamefont {Bower},
  \citenamefont {Meitl}, \citenamefont {Radauscher}, \citenamefont {Bonafede},
  \citenamefont {Pearson}, \citenamefont {Raymond}, \citenamefont {Vick},
  \citenamefont {Verreen}, \citenamefont {Weeks}, \citenamefont {Gomez},
  \citenamefont {Moore},\ and\ \citenamefont {Rotzoll}}]{xCeleprint:18}%
  \BibitemOpen
  \bibfield  {author} {\bibinfo {author} {\bibfnamefont {C.~A.}\ \bibnamefont
  {Bower}}, \bibinfo {author} {\bibfnamefont {M.}~\bibnamefont {Meitl}},
  \bibinfo {author} {\bibfnamefont {E.}~\bibnamefont {Radauscher}}, \bibinfo
  {author} {\bibfnamefont {S.}~\bibnamefont {Bonafede}}, \bibinfo {author}
  {\bibfnamefont {A.}~\bibnamefont {Pearson}}, \bibinfo {author} {\bibfnamefont
  {B.}~\bibnamefont {Raymond}}, \bibinfo {author} {\bibfnamefont
  {E.}~\bibnamefont {Vick}}, \bibinfo {author} {\bibfnamefont {C.}~\bibnamefont
  {Verreen}}, \bibinfo {author} {\bibfnamefont {T.}~\bibnamefont {Weeks}},
  \bibinfo {author} {\bibfnamefont {D.}~\bibnamefont {Gomez}}, \bibinfo
  {author} {\bibfnamefont {T.}~\bibnamefont {Moore}}, \ and\ \bibinfo {author}
  {\bibfnamefont {B.}~\bibnamefont {Rotzoll}},\ }\href@noop {} {\enquote
  {\bibinfo {title} {Printing microleds and microics for next generation
  displays},}\ }\bibinfo {howpublished}
  {https://www.xdisplay.com/wp-content/uploads/2020/05/2018\_08\_30\_IMID-updated.pdf}
  (\bibinfo {year} {2018})\BibitemShut {NoStop}%
\bibitem [{\citenamefont {{\L}ysie{\'n}}\ \emph {et~al.}(2022)\citenamefont
  {{\L}ysie{\'n}}, \citenamefont {Witczak}, \citenamefont {Wiatrowska},
  \citenamefont {Fi{\k a}czyk}, \citenamefont {Gadzali{\'n}ska}, \citenamefont
  {Schneider}, \citenamefont {Str{\k e}k}, \citenamefont {Karpi{\'n}ski},
  \citenamefont {Kosior}, \citenamefont {Granek},\ and\ \citenamefont
  {Kowalczewski}}]{Lysien:22}%
  \BibitemOpen
  \bibfield  {author} {\bibinfo {author} {\bibfnamefont {M.}~\bibnamefont
  {{\L}ysie{\'n}}}, \bibinfo {author} {\bibfnamefont {{\L}.}~\bibnamefont
  {Witczak}}, \bibinfo {author} {\bibfnamefont {A.}~\bibnamefont {Wiatrowska}},
  \bibinfo {author} {\bibfnamefont {K.}~\bibnamefont {Fi{\k a}czyk}}, \bibinfo
  {author} {\bibfnamefont {J.}~\bibnamefont {Gadzali{\'n}ska}}, \bibinfo
  {author} {\bibfnamefont {L.}~\bibnamefont {Schneider}}, \bibinfo {author}
  {\bibfnamefont {W.}~\bibnamefont {Str{\k e}k}}, \bibinfo {author}
  {\bibfnamefont {M.}~\bibnamefont {Karpi{\'n}ski}}, \bibinfo {author}
  {\bibfnamefont {{\L}.}~\bibnamefont {Kosior}}, \bibinfo {author}
  {\bibfnamefont {F.}~\bibnamefont {Granek}}, \ and\ \bibinfo {author}
  {\bibfnamefont {P.}~\bibnamefont {Kowalczewski}},\ }\href {\doibase
  10.1038/s41598-022-13352-5} {\bibfield  {journal} {\bibinfo  {journal} {Sci.
  Rep.}\ }\textbf {\bibinfo {volume} {12}},\ \bibinfo {pages} {9327} (\bibinfo
  {year} {2022})}\BibitemShut {NoStop}%
\bibitem [{\citenamefont {Ruf}\ \emph {et~al.}(2021)\citenamefont {Ruf},
  \citenamefont {Wan}, \citenamefont {Choi}, \citenamefont {Englund},\ and\
  \citenamefont {Hanson}}]{Ruf:21}%
  \BibitemOpen
  \bibfield  {author} {\bibinfo {author} {\bibfnamefont {M.}~\bibnamefont
  {Ruf}}, \bibinfo {author} {\bibfnamefont {N.~H.}\ \bibnamefont {Wan}},
  \bibinfo {author} {\bibfnamefont {H.}~\bibnamefont {Choi}}, \bibinfo {author}
  {\bibfnamefont {D.}~\bibnamefont {Englund}}, \ and\ \bibinfo {author}
  {\bibfnamefont {R.}~\bibnamefont {Hanson}},\ }\href {\doibase
  10.1063/5.0056534} {\bibfield  {journal} {\bibinfo  {journal} {J. Appl.
  Phys.}\ }\textbf {\bibinfo {volume} {130}},\ \bibinfo {pages} {070901}
  (\bibinfo {year} {2021})}\BibitemShut {NoStop}%
\bibitem [{\citenamefont {Kok}\ \emph {et~al.}(2007)\citenamefont {Kok},
  \citenamefont {Munro}, \citenamefont {Nemoto}, \citenamefont {Ralph},
  \citenamefont {Dowling},\ and\ \citenamefont {Milburn}}]{Kok:07}%
  \BibitemOpen
  \bibfield  {author} {\bibinfo {author} {\bibfnamefont {P.}~\bibnamefont
  {Kok}}, \bibinfo {author} {\bibfnamefont {W.~J.}\ \bibnamefont {Munro}},
  \bibinfo {author} {\bibfnamefont {K.}~\bibnamefont {Nemoto}}, \bibinfo
  {author} {\bibfnamefont {T.~C.}\ \bibnamefont {Ralph}}, \bibinfo {author}
  {\bibfnamefont {J.~P.}\ \bibnamefont {Dowling}}, \ and\ \bibinfo {author}
  {\bibfnamefont {G.~J.}\ \bibnamefont {Milburn}},\ }\href {\doibase
  10.1103/RevModPhys.79.135} {\bibfield  {journal} {\bibinfo  {journal} {Rev.
  Mod. Phys.}\ }\textbf {\bibinfo {volume} {79}},\ \bibinfo {pages} {135}
  (\bibinfo {year} {2007})}\BibitemShut {NoStop}%
\bibitem [{\citenamefont {Slussarenko}\ \emph {et~al.}(2017)\citenamefont
  {Slussarenko}, \citenamefont {Weston}, \citenamefont {Chrzanowski},
  \citenamefont {Shalm}, \citenamefont {Verma}, \citenamefont {Nam},\ and\
  \citenamefont {Pryde}}]{Slussarenko:17}%
  \BibitemOpen
  \bibfield  {author} {\bibinfo {author} {\bibfnamefont {S.}~\bibnamefont
  {Slussarenko}}, \bibinfo {author} {\bibfnamefont {M.~M.}\ \bibnamefont
  {Weston}}, \bibinfo {author} {\bibfnamefont {H.~M.}\ \bibnamefont
  {Chrzanowski}}, \bibinfo {author} {\bibfnamefont {L.~K.}\ \bibnamefont
  {Shalm}}, \bibinfo {author} {\bibfnamefont {V.~B.}\ \bibnamefont {Verma}},
  \bibinfo {author} {\bibfnamefont {S.~W.}\ \bibnamefont {Nam}}, \ and\
  \bibinfo {author} {\bibfnamefont {G.~J.}\ \bibnamefont {Pryde}},\ }\href
  {\doibase 10.1038/s41566-017-0011-5} {\bibfield  {journal} {\bibinfo
  {journal} {Nat. Photon.}\ }\textbf {\bibinfo {volume} {11}},\ \bibinfo
  {pages} {700} (\bibinfo {year} {2017})}\BibitemShut {NoStop}%
\bibitem [{\citenamefont {Reens}\ \emph {et~al.}(2022)\citenamefont {Reens},
  \citenamefont {Collins}, \citenamefont {Ciampi}, \citenamefont {Kharas},
  \citenamefont {Aull}, \citenamefont {Donlon}, \citenamefont {Bruzewicz},
  \citenamefont {Felton}, \citenamefont {Stuart}, \citenamefont {Niffenegger},
  \citenamefont {Rich}, \citenamefont {Braje}, \citenamefont {Ryu},
  \citenamefont {Chiaverini},\ and\ \citenamefont {McConnell}}]{Reens:22}%
  \BibitemOpen
  \bibfield  {author} {\bibinfo {author} {\bibfnamefont {D.}~\bibnamefont
  {Reens}}, \bibinfo {author} {\bibfnamefont {M.}~\bibnamefont {Collins}},
  \bibinfo {author} {\bibfnamefont {J.}~\bibnamefont {Ciampi}}, \bibinfo
  {author} {\bibfnamefont {D.}~\bibnamefont {Kharas}}, \bibinfo {author}
  {\bibfnamefont {B.~F.}\ \bibnamefont {Aull}}, \bibinfo {author}
  {\bibfnamefont {K.}~\bibnamefont {Donlon}}, \bibinfo {author} {\bibfnamefont
  {C.~D.}\ \bibnamefont {Bruzewicz}}, \bibinfo {author} {\bibfnamefont
  {B.}~\bibnamefont {Felton}}, \bibinfo {author} {\bibfnamefont
  {J.}~\bibnamefont {Stuart}}, \bibinfo {author} {\bibfnamefont {R.~J.}\
  \bibnamefont {Niffenegger}}, \bibinfo {author} {\bibfnamefont
  {P.}~\bibnamefont {Rich}}, \bibinfo {author} {\bibfnamefont {D.}~\bibnamefont
  {Braje}}, \bibinfo {author} {\bibfnamefont {K.~K.}\ \bibnamefont {Ryu}},
  \bibinfo {author} {\bibfnamefont {J.}~\bibnamefont {Chiaverini}}, \ and\
  \bibinfo {author} {\bibfnamefont {R.}~\bibnamefont {McConnell}},\ }\href
  {\doibase 10.1103/PhysRevLett.129.100502} {\bibfield  {journal} {\bibinfo
  {journal} {Phys. Rev. Lett.}\ }\textbf {\bibinfo {volume} {129}},\ \bibinfo
  {pages} {100502} (\bibinfo {year} {2022})}\BibitemShut {NoStop}%
\bibitem [{\citenamefont {Errando-Herranz}\ \emph {et~al.}(2023)\citenamefont
  {Errando-Herranz}, \citenamefont {Gyger}, \citenamefont {Tao}, \citenamefont
  {Colangelo}, \citenamefont {Christen}, \citenamefont {Larocque},
  \citenamefont {Sattari}, \citenamefont {Choong}, \citenamefont {Petremand},
  \citenamefont {Prieto}, \citenamefont {Yu}, \citenamefont {Steinhauer},
  \citenamefont {Ghadimi}, \citenamefont {Zwiller},\ and\ \citenamefont
  {Englund}}]{Errando-Herranz:23}%
  \BibitemOpen
  \bibfield  {author} {\bibinfo {author} {\bibfnamefont {C.}~\bibnamefont
  {Errando-Herranz}}, \bibinfo {author} {\bibfnamefont {S.}~\bibnamefont
  {Gyger}}, \bibinfo {author} {\bibfnamefont {M.}~\bibnamefont {Tao}}, \bibinfo
  {author} {\bibfnamefont {M.}~\bibnamefont {Colangelo}}, \bibinfo {author}
  {\bibfnamefont {I.}~\bibnamefont {Christen}}, \bibinfo {author}
  {\bibfnamefont {H.}~\bibnamefont {Larocque}}, \bibinfo {author}
  {\bibfnamefont {H.}~\bibnamefont {Sattari}}, \bibinfo {author} {\bibfnamefont
  {G.}~\bibnamefont {Choong}}, \bibinfo {author} {\bibfnamefont
  {Y.}~\bibnamefont {Petremand}}, \bibinfo {author} {\bibfnamefont
  {I.}~\bibnamefont {Prieto}}, \bibinfo {author} {\bibfnamefont
  {Y.}~\bibnamefont {Yu}}, \bibinfo {author} {\bibfnamefont {S.}~\bibnamefont
  {Steinhauer}}, \bibinfo {author} {\bibfnamefont {A.~H.}\ \bibnamefont
  {Ghadimi}}, \bibinfo {author} {\bibfnamefont {V.}~\bibnamefont {Zwiller}}, \
  and\ \bibinfo {author} {\bibfnamefont {D.}~\bibnamefont {Englund}},\ }in\
  \href {\doibase 10.1364/CLEO_FS.2023.FM2E.5} {\emph {\bibinfo {booktitle}
  {CLEO 2023}}}\ (\bibinfo  {publisher} {Optica Publishing Group},\ \bibinfo
  {year} {2023})\ p.\ \bibinfo {pages} {FM2E.5}\BibitemShut {NoStop}%
\end{thebibliography}%


\begin{thebibliography}{3}%
\makeatletter
\providecommand \@ifxundefined [1]{%
 \@ifx{#1\undefined}
}%
\providecommand \@ifnum [1]{%
 \ifnum #1\expandafter \@firstoftwo
 \else \expandafter \@secondoftwo
 \fi
}%
\providecommand \@ifx [1]{%
 \ifx #1\expandafter \@firstoftwo
 \else \expandafter \@secondoftwo
 \fi
}%
\providecommand \natexlab [1]{#1}%
\providecommand \enquote  [1]{``#1''}%
\providecommand \bibnamefont  [1]{#1}%
\providecommand \bibfnamefont [1]{#1}%
\providecommand \citenamefont [1]{#1}%
\providecommand \href@noop [0]{\@secondoftwo}%
\providecommand \href [0]{\begingroup \@sanitize@url \@href}%
\providecommand \@href[1]{\@@startlink{#1}\@@href}%
\providecommand \@@href[1]{\endgroup#1\@@endlink}%
\providecommand \@sanitize@url [0]{\catcode `\\12\catcode `\$12\catcode
  `\&12\catcode `\#12\catcode `\^12\catcode `\_12\catcode `\%12\relax}%
\providecommand \@@startlink[1]{}%
\providecommand \@@endlink[0]{}%
\providecommand \url  [0]{\begingroup\@sanitize@url \@url }%
\providecommand \@url [1]{\endgroup\@href {#1}{\urlprefix }}%
\providecommand \urlprefix  [0]{URL }%
\providecommand \Eprint [0]{\href }%
\providecommand \doibase [0]{http://dx.doi.org/}%
\providecommand \selectlanguage [0]{\@gobble}%
\providecommand \bibinfo  [0]{\@secondoftwo}%
\providecommand \bibfield  [0]{\@secondoftwo}%
\providecommand \translation [1]{[#1]}%
\providecommand \BibitemOpen [0]{}%
\providecommand \bibitemStop [0]{}%
\providecommand \bibitemNoStop [0]{.\EOS\space}%
\providecommand \EOS [0]{\spacefactor3000\relax}%
\providecommand \BibitemShut  [1]{\csname bibitem#1\endcsname}%
\let\auto@bib@innerbib\@empty
\bibitem [{\citenamefont {Timurdogan}\ \emph {et~al.}(2019)\citenamefont
  {Timurdogan}, \citenamefont {Su}, \citenamefont {Shiue}, \citenamefont
  {Poulton}, \citenamefont {Byrd}, \citenamefont {Xin},\ and\ \citenamefont
  {Watts}}]{Timurdogan:19}%
  \BibitemOpen
  \bibfield  {author} {\bibinfo {author} {\bibfnamefont {Erman}\ \bibnamefont
  {Timurdogan}}, \bibinfo {author} {\bibfnamefont {Zhan}\ \bibnamefont {Su}},
  \bibinfo {author} {\bibfnamefont {Ren-Jye}\ \bibnamefont {Shiue}}, \bibinfo
  {author} {\bibfnamefont {Christopher~V.}\ \bibnamefont {Poulton}}, \bibinfo
  {author} {\bibfnamefont {Matthew~J.}\ \bibnamefont {Byrd}}, \bibinfo {author}
  {\bibfnamefont {Simon}\ \bibnamefont {Xin}}, \ and\ \bibinfo {author}
  {\bibfnamefont {Michael~R.}\ \bibnamefont {Watts}},\ }\bibfield  {title}
  {\enquote {\bibinfo {title} {Apsuny process design kit (pdkv3.0): O, c and l
  band silicon photonics component libraries on 300mm wafers},}\ }in\ \href
  {\doibase 10.1364/OFC.2019.Tu2A.1} {\emph {\bibinfo {booktitle} {Optical
  Fiber Communication Conference (OFC) 2019}}}\ (\bibinfo  {publisher} {Optica
  Publishing Group},\ \bibinfo {year} {2019})\ p.\ \bibinfo {pages}
  {Tu2A.1}\BibitemShut {NoStop}%
\bibitem [{\citenamefont {Fahrenkopf}\ \emph {et~al.}(2019)\citenamefont
  {Fahrenkopf}, \citenamefont {McDonough}, \citenamefont {Leake}, \citenamefont
  {Su}, \citenamefont {Timurdogan},\ and\ \citenamefont
  {Coolbaugh}}]{Fahrenkopf:19}%
  \BibitemOpen
  \bibfield  {author} {\bibinfo {author} {\bibfnamefont {Nicholas~M.}\
  \bibnamefont {Fahrenkopf}}, \bibinfo {author} {\bibfnamefont {Colin}\
  \bibnamefont {McDonough}}, \bibinfo {author} {\bibfnamefont {Gerald~L.}\
  \bibnamefont {Leake}}, \bibinfo {author} {\bibfnamefont {Zhan}\ \bibnamefont
  {Su}}, \bibinfo {author} {\bibfnamefont {Erman}\ \bibnamefont {Timurdogan}},
  \ and\ \bibinfo {author} {\bibfnamefont {Douglas~D.}\ \bibnamefont
  {Coolbaugh}},\ }\bibfield  {title} {\enquote {\bibinfo {title} {The aim
  photonics mpw: A highly accessible cutting edge technology for rapid
  prototyping of photonic integrated circuits},}\ }\href {\doibase
  10.1109/JSTQE.2019.2935698} {\bibfield  {journal} {\bibinfo  {journal} {IEEE
  J. Sel. Top. Quantum Electron.}\ }\textbf {\bibinfo {volume} {25}},\ \bibinfo
  {pages} {1--6} (\bibinfo {year} {2019})}\BibitemShut {NoStop}%
\bibitem [{\citenamefont {Vandekerckhove}\ \emph {et~al.}(2023)\citenamefont
  {Vandekerckhove}, \citenamefont {Vanackere}, \citenamefont {Witte},
  \citenamefont {Cuyvers}, \citenamefont {Reis}, \citenamefont {Billet},
  \citenamefont {Roelkens}, \citenamefont {Clemmen},\ and\ \citenamefont
  {Kuyken}}]{Vandekerckhove:23}%
  \BibitemOpen
  \bibfield  {author} {\bibinfo {author} {\bibfnamefont {Tom}\ \bibnamefont
  {Vandekerckhove}}, \bibinfo {author} {\bibfnamefont {Tom}\ \bibnamefont
  {Vanackere}}, \bibinfo {author} {\bibfnamefont {Jasper~De}\ \bibnamefont
  {Witte}}, \bibinfo {author} {\bibfnamefont {Stijn}\ \bibnamefont {Cuyvers}},
  \bibinfo {author} {\bibfnamefont {Luis}\ \bibnamefont {Reis}}, \bibinfo
  {author} {\bibfnamefont {Maximilien}\ \bibnamefont {Billet}}, \bibinfo
  {author} {\bibfnamefont {G\"{u}nther}\ \bibnamefont {Roelkens}}, \bibinfo
  {author} {\bibfnamefont {St\'{e}phane}\ \bibnamefont {Clemmen}}, \ and\
  \bibinfo {author} {\bibfnamefont {Bart}\ \bibnamefont {Kuyken}},\ }\bibfield
  {title} {\enquote {\bibinfo {title} {Reliable micro-transfer printing method
  for heterogeneous integration of lithium niobate and semiconductor thin
  films},}\ }\href {\doibase 10.1364/OME.494038} {\bibfield  {journal}
  {\bibinfo  {journal} {Opt. Mater. Express}\ }\textbf {\bibinfo {volume}
  {13}},\ \bibinfo {pages} {1984--1993} (\bibinfo {year} {2023})}\BibitemShut
  {NoStop}%
\end{thebibliography}%

\end{document}


\title{Supplementary Information: Single-photon detectors on arbitrary photonic substrates}

\author{Max~Tao}
\thanks{These two authors contributed equally}
\affiliation{Research Laboratory of Electronics, Massachusetts Institute of Technology, Cambridge, Massachusetts 02139, USA}

\author{Hugo~Larocque}
\thanks{These two authors contributed equally}
\affiliation{Research Laboratory of Electronics, Massachusetts Institute of Technology, Cambridge, Massachusetts 02139, USA}

\author{Samuel~Gyger}
\affiliation{Research Laboratory of Electronics, Massachusetts Institute of Technology, Cambridge, Massachusetts 02139, USA}
\affiliation{KTH Royal Institute of Technology, Stockholm, Sweden}

\author{Marco~Colangelo}
\affiliation{Research Laboratory of Electronics, Massachusetts Institute of Technology, Cambridge, Massachusetts 02139, USA}

\author{Owen~Medeiros}
\affiliation{Research Laboratory of Electronics, Massachusetts Institute of Technology, Cambridge, Massachusetts 02139, USA}

\author{Ian~Christen}
\affiliation{Research Laboratory of Electronics, Massachusetts Institute of Technology, Cambridge, Massachusetts 02139, USA}

\author{Hamed~Sattari}
\affiliation{Centre Suisse d’Electronique et de Microtechnique, Neuchatel, Switzerland}

\author{Gregory~Choong}
\affiliation{Centre Suisse d’Electronique et de Microtechnique, Neuchatel, Switzerland}

\author{Yves~Petremand}
\affiliation{Centre Suisse d’Electronique et de Microtechnique, Neuchatel, Switzerland}

\author{Ivan~Prieto}
\affiliation{Centre Suisse d’Electronique et de Microtechnique, Neuchatel, Switzerland}

\author{Yang~Yu}
\affiliation{Raith America Inc., Troy, NY, USA}

\author{Stephan~Steinhauer}
\affiliation{KTH Royal Institute of Technology, Stockholm, Sweden}

\author{Gerald~L.~Leake}
\affiliation{State University of New York Polytechnic Institute, Albany, New York 12203, USA}

\author{Daniel~J.~Coleman}
\affiliation{State University of New York Polytechnic Institute, Albany, New York 12203, USA}

\author{Amir~H.~Ghadimi}
\affiliation{Centre Suisse d’Electronique et de Microtechnique, Neuchatel, Switzerland}

\author{Michael~L.~Fanto}
\affiliation{Air Force Research Laboratory, Information Directorate, Rome, New York, 13441, USA}

\author{Val~Zwiller}
\affiliation{KTH Royal Institute of Technology, Stockholm, Sweden}

\author{Dirk~Englund}
\affiliation{Research Laboratory of Electronics, Massachusetts Institute of Technology, Cambridge, Massachusetts 02139, USA}

\author{Carlos~Errando-Herranz}
\email{c.errandoherranz@tudelft.nl}
\affiliation{Research Laboratory of Electronics, Massachusetts Institute of Technology, Cambridge, Massachusetts 02139, USA}
\affiliation{Institute of Physics, University of M\"unster, 48149, M\"unster, Germany}
\affiliation{Department of Quantum and Computer Engineering, Delft University of Technology, Delft, Netherlands}
\affiliation{QuTech, Delft University of Technology, Delft, Netherlands}

\newpage
\onecolumngrid
\renewcommand{\thefigure}{S\arabic{figure}}
\setcounter{figure}{0}

\newpage

\maketitle

\section{SNSPD Fabrication}

\begin{figure}[htbp]
  \centering
\includegraphics[width=0.95\columnwidth]{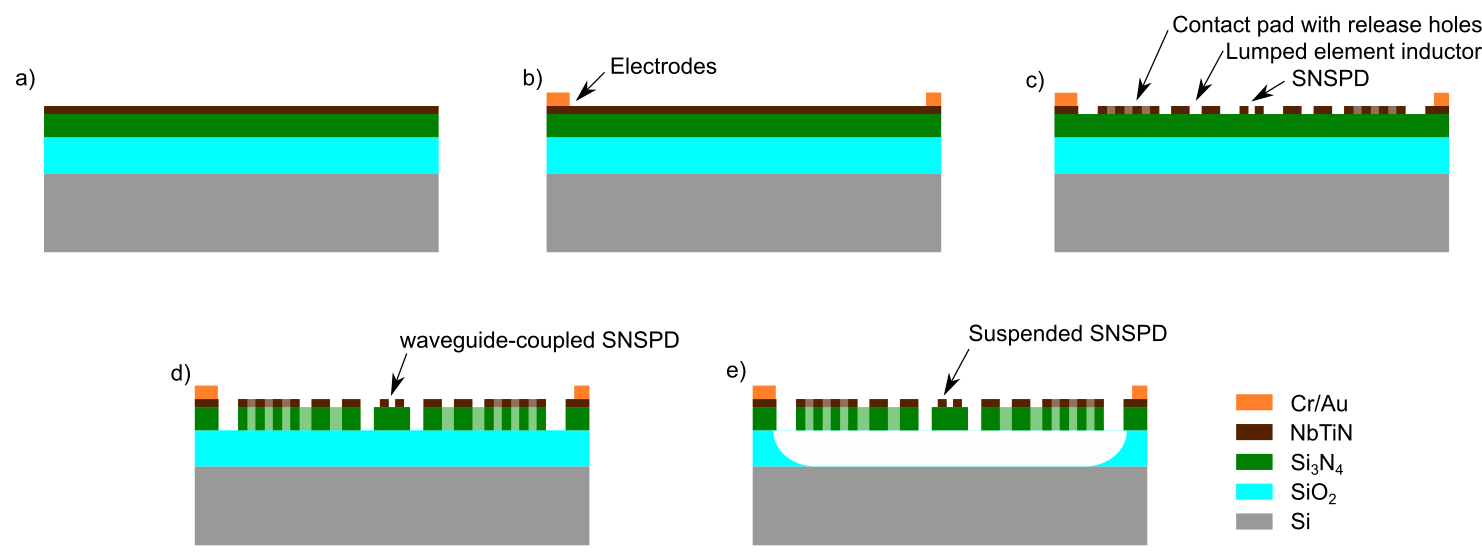}
\caption{\textbf{SNSPD Fabrication Flow.} Diagrams of the fabricated SNSPDs at various stages of their fabrication flow. }
\label{fig:fabrication_dev}
\end{figure}

Figure~\ref{fig:fabrication_dev} shows the overall fabrication process flow for the SNSPD chiplets. We fabricate the waveguide-coupled SNSPDs from a foundry silicon nitride on insulator wafer with 250~nm of stoichiometric LPCVD silicon nitride and 2~\textmu m of buried oxide. First, we deposit 9~nm of NbTiN on the wafer and then pattern nanostructures using electron beam lithography followed by a directional reactive ion etch. The patterned superconducting film includes a 90~nm wide hairpin nanowire detector connected to a lumped element inductor and contact pads with release holes for under-etching. We then pattern the silicon nitride to produce the device's waveguide and release holes. The waveguide width is 1~\textmu m and tapers down to 100~nm over a 10~\textmu m length. 
Beyond this tapered region, the detector waveguide preserves its nominal 1~\textmu m width over 90~\textmu m. 
This length completely overlaps with the extent of the structure's hairpin detector. Tethers connect these waveguides to pads with release holes serving as the main point of contact with the elastomer stamp during transfer printing while providing the required adhesion with the PIC during its placement. We finally release the resulting structure using a buffered oxide etch followed by critical point drying to remove the buried oxide below the nitride film. After the full fabrication process, we measure the resistance across the nanowire to ensure they are still intact and screen for faulty devices showing an open circuit.

\section{Hybrid Structure Fabrication}
\begin{figure}[htbp]
  \centering
\includegraphics[width=0.95\columnwidth]{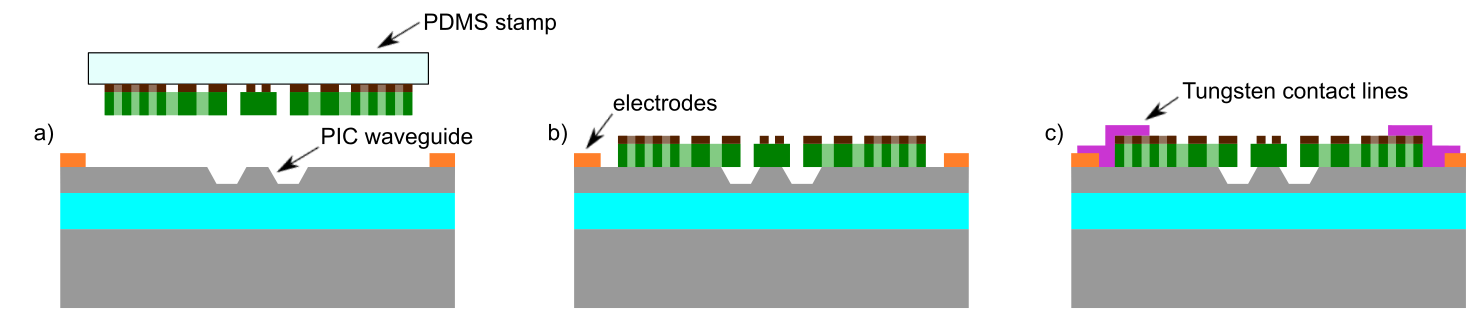}
\caption{\textbf{Hybrid Integration Fabrication Flow} Diagrams of the hybrid SNSPD - PIC structure at various stages of the fabrication flow.}
\label{fig:fabrication_hybrid}
\end{figure}

The process flow for the fabrication of the hybrid structure follows Fig.~\ref{fig:fabrication_hybrid}. For the case of the commercial silicon PICs, we first pattern gold wirebonding pads around the PIC waveguide to which we can electrically connect the SNSPDs for testing. Specifically, we rely on a post-foundry fabrication scheme involving optical lithography, electron-beam physical vapor deposition of 10~nm chromium and 40~nm of gold, followed by an overnight liftoff process. This process is not necessary for our integration method, and was only used to make the FIB-deposited wiring shorter and thus faster in time. For the LNOI chip we used the pre-existing foundry metals without any post-process metalization step.

Using a 50~\textmu m~$\times$~50~\textmu m~$\times$~50~\textmu m PDMS stamp, we then transfer the released devices to the photonic chip while optically aligning the waveguide-coupled SNSPD to the corresponding waveguide on the PIC. 

To wire our SNSPDs' superconducting film to our chips' contact pads, we deposited tungsten lines using a Raith VELION FIB-SEM system with an Au+ beam set to a 35~kV acceleration voltage, a 120 pA current, and a dose of 2~nC/~\textmu m$^2$. We set the system to deposit lines that were 1~\textmu m wide and 5~\textmu m long. Our ion beam settings resulted in 500~nm thick lines with residual sputtering from the beam tail up to 2~\textmu m from the targeted deposition sites.

\section{Experimental Setup}

\begin{figure}[htbp]
  \centering
  \includegraphics[width=\columnwidth]{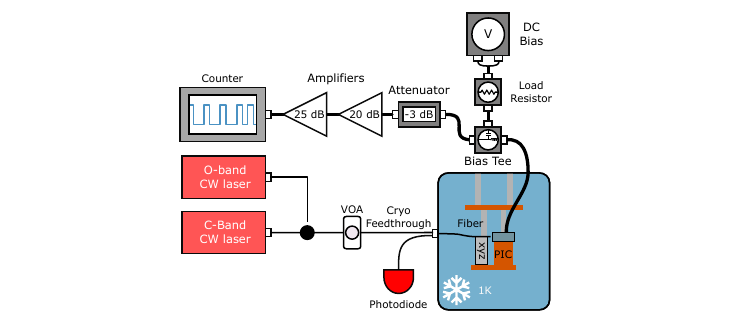}
\caption{\textbf{Apparatus for cryogenic testing.} Figure legend: VOA: Variable Optical Attenuator, PIC: Photonic integrated circuit, xyz: xyz piezo-electrically actuated positioner.}
\label{fig:setup}
\end{figure}

Figure~\ref{fig:setup} illustrates the schematics for the apparatus used to characterize our hybrid integrated SNSPDs at cryogenic temperatures. For testing the detectors on silicon PICs, we mount our hybrid chips in an ICEoxford 1K cryostat, where a UHNA1 optical fiber array mounted on a 3-axis Attocube piezo positioner stack provides the required optical input/output to the circuit by aligning it to its edge couplers~\cite{Timurdogan:19}. Splices to SMF28 fibers fed through the cryostat ensure optical connectivity to components outside the chamber. These components notably include continuous wave O-band and C-band O-band tunable external cavity diode lasers (Santec TSL-570 and Santec TSL-710), which consisted of the main source of photons for our experiments, along with a variable optical attenuator (JDS HJA9) limiting their output power. While testing the lithium niobate hybrid PICs operating at visible wavelengths, we replaced these tunable sources with a fiber-coupled laser source emitting light near a 650~nm wavelength (Thorlabs S4FC series). We glued the chip to a custom-machined copper plate with thermally-conductive glue and wire bonded it to a printed circuit board. We perform room temperature alignment of the fiber array to the PIC edge couplers first by monitoring optical power on a photodiode (Thorlabs S122C) through a loop-back structure on the chip. A feed-back loop optimizing the strength of this signal then preserves the alignment of the fiber array relative to the PIC while cooling down the chamber. To monitor single photon counts coming from the SNSPDs, we biased them with a DC current using a combination of a bias tee (Mini-Circuits ZFBT-4R2G+), a load resistor (100 k$\Omega$), and voltage source (SRS SIM928). A sequence of components consisting of an attenuator and low-noise amplifiers (RF Bay LNA-2500, LNA-2000) amplified the acquired biased signal before sending it to a counter (Agilent 53131a).

\section{Detector Efficiency Modeling}

Low overlap between the hybrid mode and the SNSPD over most of the chiplet's length could be responsible for the low optical detection efficiencies (ODE) in the hybrid silicon PICs. Here, the PIC waveguides specifically consists of 220~nm-thick and 400~nm-wide silicon waveguides designed for single-mode O- and C+L-band operation. Underneath the transferred silicon nitride waveguides, their widths taper down to 200~nm over a length of 40~\textmu m. 

\begin{figure}[htbp]
  \centering
  \includegraphics[width=\columnwidth]{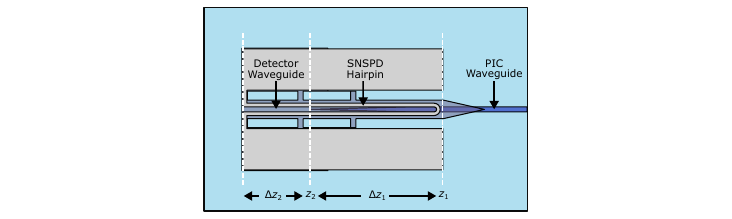}
\caption{\textbf{Parameters related to the hybrid detector efficiency}. $z_1$ and $z_2$ correspond to the start of the hairpin diode and to the end of the PIC waveguide, respectively. We separate the length of the hairpin among $\Delta z_1 = z_2-z_1$ and $\Delta z_2$.}
\label{fig:zGeometry}
\end{figure}

When the constituent waveguides of the detector's hybrid mode converter are perfectly aligned, we expect the converter's fundamental mode to mostly contain light coming from the PIC's waveguide. We expect the evolution of this mode, $A(x,y)$, to verify
\begin{eqnarray}
    \frac{\partial A}{\partial z} &=& (-i k(z) + \alpha (z)) A(z), \\
    \label{eq:PDE}
   A(z,\Delta z) &\propto& \exp\left(\int_{z'=z}^{z'=z+\Delta z} -i k(z')+ \alpha(z') \, dz'\right),
\end{eqnarray}
where $A(z)$ is the weight of the mode as it propagates along $z$, while $k(z)$ and $\alpha(z)$ are the $z$-dependent real and imaginary components of the mode's propagation constants. The ODE of the device thus becomes $1-|A(L)|^2$, where $L$ is the length of the hybrid structure.
Losses incurred at abrupt changes in the hybrid mode converter can further alter this metric. We expect these losses to primarily arise at the tip of the mode converter's tapers. Assuming that losses at the tip of the detector and PIC waveguide tapers are given by $t_\text{det}$ and $t_\text{PIC}$, respectively, then the expected ODE of our hybrid detectors corresponds to 
\begin{equation}
    \label{eq:odeFEM}
    |t_\text{det}|^2 (1-|A(z_1,\Delta z_1)|^2) + |t_\text{det}|^2 |t_\text{PIC}|^2 |A(z_1,\Delta z_1)|^2 (1-|A(z_2,\Delta z_2)|^2)
\end{equation}
where we schematically define the positions $z_{1,2}$ and intervals $\Delta z_{1,2}$ in Fig.~\ref{fig:zGeometry}.
From finite difference time domain (FDTD) simulations, we estimate $|t_\text{det}|^2=0.995$ and $|t_\text{PIC}|^2=0.925$. To evaluate the $A(z, \Delta z)$ terms, we rely on numerically calculated values of the effective refractive indices of the fundamental mode along the mode converter using a finite element method (FEM). From these indices, we extract the propagation constants $k(z)$ and $\alpha(z)$, thus allowing us to compute $A(z, \Delta z)$ using Eq.~(\ref{eq:PDE}). From these simulations, Eq.~(\ref{eq:odeFEM}) suggests an expected detection efficiency of 30.3\% for a perfectly transferred SNSPD on our silicon PIC operating near a wavelength of 1570~nm. 

\begin{figure}[htbp]
  \centering
  \includegraphics[width=\columnwidth]{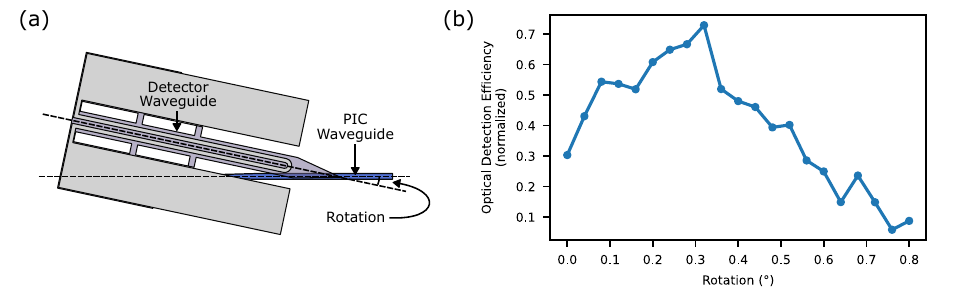}
\caption{\textbf{Expected optical detection efficiency for hybrid detectors}. (a) Diagram illustrating the observed rotational offset incurred during the transfer of our detectors on silicon PICs. (b) We consider detectors transferred on a silicon PIC operating at an optical wavelength of 1570~nm and featuring various rotational offsets between their constituent waveguides.}
\label{fig:efficiencyRotation}
\end{figure}

When the transferred detector exhibits a rotation offset from its perfectly aligned configuration as sketched out in Fig.~\ref{fig:efficiencyRotation}(a), the resulting asymmetries lead to the excitation of higher order modes in the hybrid mode converter. The resulting coupling between the device's eigenmodes therefore causes Eqs.~(\ref{eq:PDE},\ref{eq:odeFEM}) to lose their validity. To capture such effects, we numerically propagate the fundamental TE mode of the silicon PIC waveguide through the mode converter using an FEM method implemented in RSoft Photonic Device Tools. The software provides the energy absorbed by the device, which it calculates using the following integral,
\begin{equation}
    U_A = \omega \int_V \text{Im} [ \epsilon(\omega) ] |\mathbf{E}(\omega)|^2 \, dV
\end{equation}
where $\omega$ is the optical frequency, $\epsilon$ is the dielectric permittivity and $\mathbf{E}$ is the electric field. For a straight waveguide, $U_A$ is proportional to the $1-|A(L)|^2$ expression used to calculate efficiency. To estimate the influence of misalignment on our ODE, we therefore normalize these absorbed energy values to the 30.3\% ODE expected from the effective indices of the perfectly aligned device. Figure~\ref{fig:efficiencyRotation}(b) provides these normalized efficiency values for hybrid structures with detector waveguides exhibiting various rotational offsets with respect to the PIC waveguide.
Optical imaging of the hybrid device considered in this work indicates a $0.8^\text{o}$ offset between the two waveguides. From Fig.~\ref{fig:efficiencyRotation}(b), we expect this offset to reduce our ODE near 8.7\%, which is near our $7.8 \pm 0.2\%$ measured value. We attribute the increased ODE at intermediate rotation angles to the excitation of the hybrid structure's TM0 mode, which features greater overlap with the device's SNSPD, thereby leading to enhanced optical absorption.

Co-designing the PIC and detector waveguides in the hybrid device would increase its detection efficiency. Given the constraint of working with the considered 250~nm $\times$ 1~\textmu m silicon nitride detector waveguides integrated on a silicon photonic PIC, such co-design could involve using a silicon nitride PIC waveguide, which are increasingly prevalent in commercial silicon PIC manufacturing processes~\cite{Fahrenkopf:19}. For instance, at the considered 1570~nm optical wavelength, replacing our PIC's silicon taper by a 220~nm thick silicon nitride waveguide that tapers from a width of 1~\textmu m down to 200~nm over a length of 20~\textmu m increased the ODE near 80~\%. At this point, the overlap between the detector waveguide mode and the SNSPD primarily limits detection efficiency. Besides increasing this overlap with a modified detector waveguide geometry, increasing the length of this waveguide provides another venue for increasing the ODE. For example, we estimate that a detector waveguide length of 250~\textmu m ought to increase efficiency above 99\%. Such longer detectors would be compatible with transfer printing, as this integration method has been shown to accommodate longer photonic structures exceeding 2~mm~\cite{Vandekerckhove:23}.

\section{SNSPD Characterization}

\begin{figure}[htbp]
  \centering
  \includegraphics[width=\columnwidth]{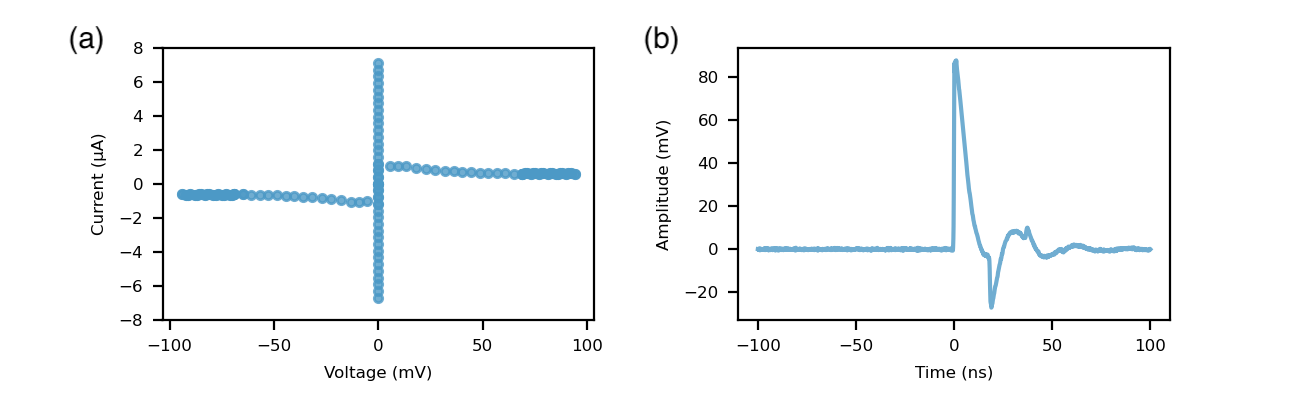}
  \caption{\textbf{Electrical response of hybrid integrated SNSPDs at cryogenic temperatures of 1~K.} (a) IV curve of the SNSPD. (b) Pulse profile of the SNSPD following the detection of a photon. }
\label{fig:electrical}
\end{figure}

Figure~\ref{fig:electrical} shows the electrical response of our hybrid integrated SNSPDs to various bias currents and to a detected single photon. Figure~\ref{fig:electrical}(a) plots the IV curve of the device, thereby indicating a switching current of 7.1~\textmu A. Figure~\ref{fig:electrical}(b) plots the pulse profile of the detector attributed to the detection of a photon, which features a decay time of $\sim 50$~ns along with secondary peaks likely attributed to reflections from the amplifiers shown in Fig.~\ref{fig:setup}.

\begin{figure}[htbp]
  \centering
\includegraphics[width=\columnwidth]{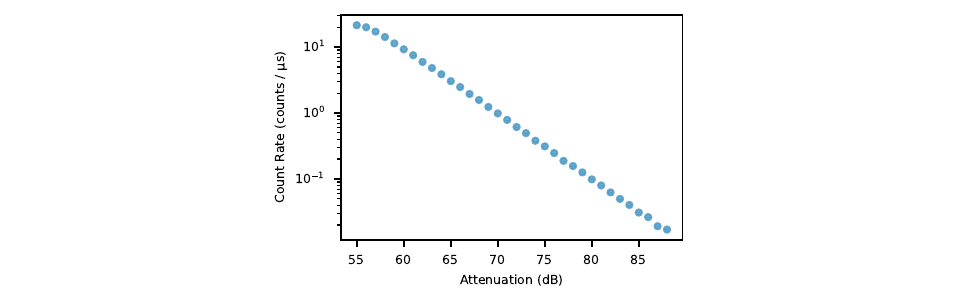}
\caption{\textbf{SNSPD count-rate dependence on incident photon flux.} SNSPD count-rate vs the attenuation of our input optical signal.}
\label{fig:linearity}
\end{figure}

\begin{figure}[htbp]
  \centering
  \includegraphics[width=\columnwidth]{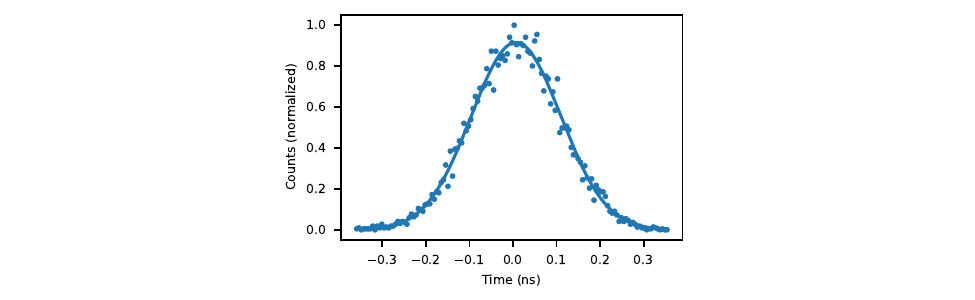}
  \caption{\textbf{Jitter Measurement.} System jitter measurement indicating a detector jitter of 242~ps. }
\label{fig:jitter}
\end{figure}

To verify the linear response of our detector's counts to the number of incident photons, we monitor this quantity while decreasing our input's optical power with our variable optical attenuator. Figure~\ref{fig:linearity} plots the results of this measurement and features a distinct linear drop in count-rate as we increase the attenuation of our input. Finally, as shown in Fig.~\ref{fig:jitter}, we measure a 242~ps jitter for our integrated detectors.

\section{Off-Chip Optical Transmission Calibration}

\begin{figure}[htbp]
    \centering
    \includegraphics[width=\columnwidth]{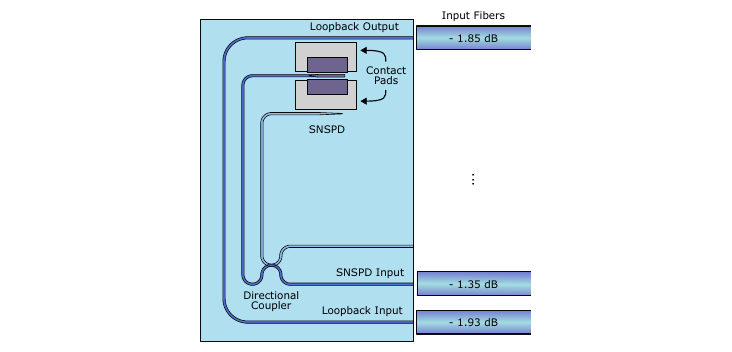}
    \caption{\textbf{Schematic of silicon hybrid PIC.} Our testing apparatus relies on three input fibers going through the cryo feedthrough and exhibiting different insertion losses. Two of them couple to an on-chip loopback structure allowing us to calibrate facet losses from the PIC. A directional coupler lies between the last facet and the SNSPD. To minimize its contribution to the measurement, we alter the optical wavelength used in our experiments such that most of the input light gets routed to the coupler's bar port. The schematics also provides the insertion loss of each fiber due to the splices to the cryo feedthrough.}
    \label{fig:pic_schematic}
\end{figure}

To estimate the optical power incident on our hybrid integrated detectors, we calibrated off-chip transmission losses through the various optical components inserted between our lasers and our PIC's waveguides. We summarize these values for the apparatus used to characterize our silicon PICs in Table~\ref{tab:losses}. We measured losses due to the cryo feedthrough, i.e. the ``cryo feedthrough to fiber facet'' value, by measuring optical power coming out of its fiber array while sending light through the spliced input SMF28 fibers. As emphasized in Fig.~\ref{fig:pic_schematic}, the value provided in Table~\ref{tab:losses} corresponds to losses for the fiber coupled to the waveguide leading to the characterized SNSPD. Furthermore, we estimate the average PIC facet loss by measuring transmission through our chip's loopback structure as shown in Fig.~\ref{fig:pic_schematic}, removing the contribution of the feedthrough losses of the employed fibers, and assuming that the losses at the two PIC facets equally contribute to the resulting metric. In addition to losses due to components before the feedthrough, the feedthrough itself, and the PIC facet, we further weaken our input signal with the variable optical attenuator shown in Fig.~\ref{fig:setup}. As indicated in Table~\ref{tab:losses}, we set its attenuation to 70~dB while characterizing our SNSPD. 

\begin{table}
    \centering
    \begin{tabular}{|p{6cm}|p{2cm}|p{2cm}|} 
    \hline
    \rowcolor{Gray}
      \thead{\bf Optical Path Stage}  & \thead{\bf 1570 nm} & \thead{\bf 1312 nm} \\ \hline 
        \makecell{Laser to cryo feedthrough} & \makecell{4.18 dB} & \makecell{4.65 dB} \\ \hline \rowcolor{Gray}
        \makecell{Cryo feedthrough to fiber facet} & \makecell{1.35 dB} & \makecell{1.35 dB} \\ \hline
        \makecell{Average PIC facet loss} & \makecell{8.35 dB} & \makecell{11.05 dB} \\ \hline \rowcolor{Gray}
        \makecell{Attenuator setting during SNSPD testing} & \makecell{70~dB} & \makecell{70~dB}\\ \hline 
    \end{tabular}
    \caption{{\textbf{Optical transmission losses of the fiber optics components placed before the PIC.}} Transmission losses of the optical components placed between the hybrid Si PIC and the laser used to characterize them.}
    \label{tab:losses}
\end{table}

We assume that the losses listed in Table~\ref{tab:losses} completely account for the attenuation of the incident optical signal at the hybrid integrated detectors. As illustrated in Fig.~\ref{fig:pic_schematic}, unaccounted additional losses could arise from a directional coupler placed between the facet and the SNSPD. To mitigate these losses, we varied the optical wavelength of our input light until the counts on our SNSPD were maximized, thereby suggesting that the coupler routed most of the input light into its bar port. As indicated in Table~\ref{tab:losses} and in the main text, these optimal wavelength consist of 1570~nm for the C+L band and 1312~nm for the O-band.

\section{On-chip detection efficiency error}

We calculate the error on our reported ODE values based on the error on our measured counts and that of the optical transmission through the fiber array feed-through coupling light into the PIC, $\text{dB}_\text{f}$. Because we extracted the value by manually holding the fiber array in front of a power meter, we expect the resulting fluctuations to dominate the error on the expected photon flux hitting the detector, as defined in Eq.~(1) of the main text. We extract the error on the measured counts from the standard deviation of the seven points in Fig.~2 of the main text nearest to the one acquired at the reported bias currents. We then use standard propagation of uncertainty methods to extract the error on our ODE defined as $\Phi_\text{measured}/\Phi_\text{expected}$.

\bibliography{MEMSsnspd}